\documentstyle[12pt]{article}

\begin{document}

\baselineskip = 12pt

\def\la{\mathrel{\mathpalette\fun <}}
\def\ga{\mathrel{\mathpalette\fun >}}
\def\fun#1#2{\lower3.6pt\vbox{\baselineskip0pt\lineskip.9pt
  \ialign{$\mathsurround=0pt#1\hfil##\hfil$\crcr#2\crcr\sim\crcr}}}
%less than approximately and greater than approximately

\def\xvec{{\bf x}}
\def\kvec{{\bf k}}
\def\nvec{{\bf n}}
\def\qvec{{\bf q}}
\def\delt2{\langle \delta^2 \rangle}

\hskip 4in OSU-TA-7/96

\vskip 1.25in

\centerline{\bf LOCAL LAGRANGIAN APPROXIMATIONS FOR THE EVOLUTION OF THE}

\centerline{\bf DENSITY DISTRIBUTION FUNCTION IN LARGE-SCALE STRUCTURE}

\vskip .2in

\centerline{Zacharias A.M. Protogeros and Robert J. Scherrer}

\centerline{Department of Physics}

\centerline{The Ohio State University}

\centerline{Columbus, OH 43210}

\vskip 1in

\centerline{ABSTRACT}

We examine local Lagrangian approximations for the gravitational evolution of the density
distribution function.  In these approximations, the final density at a Lagrangian
point $\qvec$ at a time $t$ is taken to be a function only of $t$ and of the initial density
at the same Lagrangian point.  A general expression is given for the evolved density
distribution function for such approximations, and we show that the vertex generating
function for a local Lagrangian mapping applied to an initially Gaussian density field
bears a simple relation to the mapping itself.
Using this result, we design a local Lagrangian mapping which reproduces nearly exactly
the hierarchical
amplitudes given by perturbation theory for gravitational evolution.  When
extended to smoothed density fields and
applied to Gaussian initial conditions, this mapping produces a final density
distribution
function in excellent agreement with full numerical simulations of gravitational
clustering.  We also examine the application of these local Lagrangian approximations
to non-Gaussian initial conditions. 

\vskip 1 cm
\vfill
\eject
\noindent{\bf 1 INTRODUCTION}

\indent

A variety of statistics have been developed to describe the evolution of
the large-scale matter distribution in the universe.  Among these
is the one-point probability distribution function (PDF) of the density
field, $P(\rho)$, which gives the probability that the density at a random
point in space lies between $\rho$ and $\rho + d\rho$.  In the linear regime,
the entire density field is simply scaled up the growth factor $D(t)$, so
$P(\rho)$ (with $\rho$ suitably rescaled) remains constant.  In the nonlinear
regime, however, $P(\rho)$ evolves in a complex manner.
A number of recent
studies have examined the evolution of the PDF of the density field
during gravitational clustering in the nonlinear regime.
Kofman (1991), Kofman et al. (1994), and Bernardeau \& Kofman (1995)
used the Zel'dovich approximation
and the condition of mass conservation to derive an approximate
expression for $P(\rho)$.  Padmanabhan \& Subramanian (1993) derived an
approximation to the smoothed PDF from the Zel'dovich approximation.
Juszkiewicz et al. (1994) used the
Edgeworth expansion along with the moments of the evolved distribution
to obtain an approximation for the evolved PDF (see also Bernardeau \& Kofman 1995.)
[For a recent review of approximation methods in general, see Sahni \& Coles 1995].

All of these approximations assume Gaussian initial conditions; much less
is known about the evolution of the PDF for non-Gaussian initial conditions.
A number of numerical simulations have been performed to investigate
the evolution
of the density field in various non-Gaussian toy models
(Messina et al. 1990; Moscardini et al. 1991;
Matarrese et al. 1991; Weinberg \& Cole 1992; Coles et al. 1993).
Fry \& Scherrer (1994) examined analytically the evolution of skewness
in arbitrary non-Gaussian models, and this analysis was extended to
the kurtosis by Chodorowski \& Bouchet (1996),
but very little analytic work has been done
on the evolution of the full PDF in arbitrary non-Gaussian models.

In this paper, we examine a class of approximations for the evolution of the PDF
which we call ``local Lagrangian approximations".
By ``local Lagrangian", we mean that the density
at the Lagrangian point $\qvec$ at a time $t$ is approximated as a function only of
$t$ and the initial value of $\rho(\qvec)$:
\begin{equation}
\label{mapping}
\rho(\qvec,t) = f(\rho_0(\qvec), t),
\end{equation}
where $\rho_0(\qvec) \equiv \rho(\qvec,t_0)$.  [We caution the reader
that the term ``local" has been used in this context with a variety of
different meanings].
The linear approximation, for example, is a local Eulerian mapping.
The Zel'dovich approximation
is a Lagrangian mapping in which the density at $\qvec$ is a function
only of $t$ and the second partial derivatives of the initial
potential $\phi_0(\qvec)$:
\begin{equation}
\rho(\qvec,t) = f(\phi_{0_{ij}},t)
\end{equation}
For certain special cases (1-dimensional collapse, spherical collapse),
the Zel'dovich approximation reduces to a local Lagrangian
mapping of the initial density field as given in equation (\ref{mapping}).
In fact, these cases form the basis of two of our approximations.
Note
that we do not require a prescription for the Lagrangian
mapping $\xvec(\qvec)$; all we care about is the function given in equation
(\ref{mapping}),
which is sufficient to calculate the PDF $P(\rho)$.  In fact, for a given mapping
$f$ in equation (\ref{mapping}), there may not even be a Lagrangian mapping scheme which
produces $f$; nonetheless our mapping could still provide a good approximation
to the evolution of the PDF.

Our motivation for considering such mappings is two-fold:  for the case
of Gaussian initial conditions, it would be extremely interesting if a simple
mapping of the form given in equation (\ref{mapping}) could give an accurate
description of the evolution of the density PDF.  We will see that this
is indeed the case.  For non-Gaussian initial conditions, such a mapping
would allow for the investigation of the evolution of the density PDF for
a wide range of initial conditions, and it might help to answer
some general questions about the evolution of such density fields.
For example, in the quasi-linear regime,
is there any general difference between the rate at which $\langle \delta^2 \rangle$
evolves in non-Gaussian models, versus its evolution in Gaussian models?
Fry \& Scherrer (1994) suggested that in the quasi-linear regime,
models with positive initial skewness would have an rms density fluctuation
slightly larger than that predicted by linear theory, while negative skewness
models would have an rms fluctuation slightly smaller than the linear prediction.
The results of
Weinberg \& Cole (1992) support this conclusion with regard to models
with negative initial skewness but are inconclusive with regard to
models with positive initial skewness.  A second issue addressed by Fry \& Scherrer
was the evolution of skewness in non-Gaussian models.  Their results suggest
that the skewness of the evolved density field is sensitive to the initial kurtosis (as well as the initial skewness), but the expressions they derive contain integrals over the initial
three- and four-point functions, so it is difficult to draw general conclusions about the
evolution of the skewness in such models.
These are some of the many important general questions
about the evolution of non-Gaussian models which are at present unanswered.

In this paper, we consider three different models for the Lagrangian density mapping
in eq. (\ref{mapping}).  The first of these corresponds to the exact evolution in the 1-dimensional
case, and it would be exact in the case of 1-D symmetry.  Our second
model is based on the approximation method given by
Padmanabhan \& Subramanian (1993), and it corresponds to the
Zel'dovich approximation
for spherically-symmetric collapse.  We believe that the correct PDF will
lie somewhere in between these two cases.  The third approximation, in fact,
does lie between these two extremes and is constructed
to give approximately the correct hierarchical amplitudes in the quasi-linear regime.

In the next section, we present the motivation for our three approximations and 
derive the form of the density PDF for local Lagrangian mappings.
In Section 3, we use our approximations to derive
analytic perturbative results in the quasi-linear regime
for the evolution of both Gaussian and non-Gaussian initial conditions,
and we discuss the effects of smoothing.
In Section 4, we use our approximations in the nonlinear regime to derive the evolved PDF for Gaussian
initial conditions, which we compare with numerical simulations.
We then apply our approximations
to a variety of non-Gaussian initial conditions.
Our results and conclusions are summarized in Section 5.  In the
Appendix, we derive the relation between any local Lagrangian
mapping of an initially Gaussian density field
and the corresponding vertex generating function.

\vskip 1 cm
\noindent {\bf 2 LOCAL LAGRANGIAN APPROXIMATIONS}
\indent

To obtain the form of the evolved density PDF one has to use some 
approximation scheme describing the particle dynamics, as well as to adopt 
a set of initial conditions on the density field. One of the most efficient 
approximation schemes is the Zel'dovich approximation (ZA) (Zel'dovich 1970) 
which can be thought of as an operator acting on the initial (Lagrangian) 
comoving position ${\bf q}$ of a particle and yielding its final (Eulerian) 
comoving position ${\bf x}$. Specifically, one has:
\begin{equation}
\label{xq}
{\bf x}({\bf q},t) = {\bf q} + D(t){\bf \Psi}({\bf q}),
\end{equation}
where $D(t)$ is a universal time dependent function proportional to the 
expansion factor $a(t)$ for a flat, pressureless universe, 
and ${\bf \Psi}({\bf q})$ is proportional to the gradient of the initial
gravitational potential.

The evolution of the PDF of the density field in the Zel'dovich approximation
has been discussed in detail by Kofman et al. (1994); here we briefly summarize
their results relevant to our work.
Consider a particle of mass $dm$ uniformly spread at time $t_0$ inside an 
infinitesimal Lagrangian volume $d^3q$. At some later time $t$ the 
respective volume has evolved to the Eulerian infinitesimal one $d^3x_q$. 
Mass conservation implies:
\begin{equation}
\label{volume}
\rho({\bf q},t)d^3x_q = \rho({\bf q},t_0) d^3q,
\end{equation}
In the quasi-linear regime prior to shell-crossing, multi-streaming may
be neglected; here we make the assumption that multi-streaming is unimportant.

From equation (\ref{volume}) 
one obtains:
\begin{equation}
{\rho({\bf q},t)} = {\bar\rho\Vert{{\partial{\bf x}} 
\over{\partial{\bf q}}}\Vert^{-1}},
\end{equation}
According to equation (\ref{xq}) one has then:
\begin{equation}
\Vert{{\partial{\bf x}} \over{\partial{\bf q}}}\Vert^{-1} = \Vert {I + 
{D{{\partial{\bf \Psi}} \over{\partial{\bf q}}}}}\Vert^{-1}.
\end{equation}
Taking $\eta \equiv \rho/\bar \rho$,
equation (2.3) yields:
\begin{equation}
\eta = {1\over{\Vert 1 - D {{\partial{\bf \Psi}} \over{\partial{\bf q}}}\Vert}}.
\end{equation}
If $\partial \Psi_i/ \partial q_j$ has eigenvalues $\lambda_1$, $\lambda_2$,
and $\lambda_3$, then the expression for the evolved density in the Zel'dovich
approximation can be written as:
\begin{equation}
\label{zeldovich}
\eta(\qvec,t) = {\eta_0(\qvec)\over [1-D(t)\lambda_1][1-D(t)\lambda_2][1-D(t)\lambda_3]}.
\end{equation}
The distribution of $\lambda$ can be calculated exactly in the case of Gaussian initial
conditions, and the results applied to determine the exact form for $P(\rho)$ (Kofman
et al. 1994).  Unfortunately, the exact distribution of the eigenvalues
of $\partial \Psi_i/ \partial q_j$ is not easy to derive for most non-Gaussian
models, so we consider several possible ways to simplify eq. (\ref{zeldovich}).

We look for approximations to equation (\ref{zeldovich}) which are ``local",
i.e., the right hand side is a function only of $D(t)$ and $\delta_0$.
To derive approximations of this sort, we note that
the relation between $\delta_0$ and the $\lambda's$ is
\begin{equation}
\label{lambda}
\delta_0 = \lambda_1 + \lambda_2 + \lambda_3
\end{equation}
Consider first the approximation:
\begin{equation}
\eta(\qvec,t) = {\eta_0(\qvec)\over [1-D(t)(\lambda_1+\lambda_2+\lambda_3)]}.
\end{equation}
With equation (\ref{lambda}), this reduces
to
\begin{equation}
\label{planar}
\eta(\qvec,t) = {\eta_0(\qvec)\over [1-D(t)\delta_0(\qvec)]}.
\end{equation}
This approximation is exact in the limit of 1-dimensional collapse
($\lambda_2 = \lambda_3 = 0$) and corresponds to exact gravitational
evolution in one dimension (Shandarin \& Zel'dovich 1989).  Hence the results
we derive for this approximation will give the exact evolution of the PDF in one
dimension.  Equation (\ref{planar}) has also been investigated as an approximation
to the evolution of the density field in three dimensions (Nusser et al. 1991;
Mataresse et al. 1992).  We will refer to equation (\ref{planar}) as the
{\it planar} approximation.

The other extreme case is
spherical collapse, for which $\lambda_1 = \lambda_2 = \lambda_3$.
In this case, equation (\ref{zeldovich}) becomes
\begin{equation}
\label{spherical}
\eta(\qvec,t) = {\eta_0(\qvec)\over [1-D(t)\delta_0(\qvec)/3]^3}.
\end{equation}
Equation (\ref{spherical}) corresponds to the approximation of Padmanabhan and Subramanian (1993)
in the limit of zero smoothing; this approximation was also used by
Betancort-Rijo (1991) in a study of the evolution of the rms density fluctuation.
We shall refer to this as the {\it spherical} approximation.

We expect that the actual evolution of the density field lies
somewhere in between spherical and planar collapse.
Note that both equations (\ref{planar}) and (\ref{spherical}) are of the form
\begin{equation}
\label{general}
\eta(\qvec,t) = {\eta_0(\qvec) \over (1-D(t) \delta_0(\qvec)/\alpha)^\alpha},
\end{equation}
with $\alpha = 1$ for the planar approximation and $\alpha = 3$ for
the spherical approximation.  For our third approximation, we choose
$\alpha = 3/2$ to obtain:
\begin{equation}
\label{1.5}
\eta(\qvec,t) = {\eta_0(\qvec) \over [1-2 D(t) \delta_0(\qvec)/3]^{3/2}}.
\end {equation}
The choice of $\alpha = 3/2$
has no particular physical significance, but we show in the next section
[based on earlier results of Bernardeau (1992) and Bernardeau \& Kofman (1995)]
that the hierarichal amplitudes
for this model closely mimic the results of exact perturbation theory.
For that reason, we will refer to
equation (\ref{1.5}) by the oxymoronic name of the {\it exact} approximation.

How do we get from the Lagrangian mappings given by equations (\ref{planar}),
(\ref{spherical}), and (\ref{1.5}) to the
evolved density distribution function?  Following Kofman et al. (1994), we
define $P(\rho)$ (or, equivalently, $P(\eta)$) to be the Eulerian density distribution
function and take $Q(\rho)$ to be the Lagrangian distribution function.  Basically,
$P(\rho)$ gives the probability that a randomly-selected point in space has a density
in the interval $\rho$ to $\rho + d\rho$, while $Q(\rho)$ is the probability that
a randomly-selected mass point has a density in that interval.  Since the probability
of ``randomly" selecting a given mass point at an Eulerian location $\xvec$ is
proportional to the density at $\xvec$, we have (Kofman et al. 1994)
$Q(\rho) = (\rho/\bar \rho) P(\rho)$, or
\begin{equation}
\label{PQ}
P(\eta) = {1\over \eta} Q(\eta).
\end{equation}
Furthermore, we assume an initial Eulerian distribution $P_0(\eta_0)$ and initial
Lagrangian distribution $Q_0(\eta_0)$, also related by
$P_0(\eta_0) = Q_0(\eta_0)/\eta_0$.
Given a Lagrangian mapping of the form $\eta(\qvec,t) = f(\eta_0(\qvec),t)$, the evolved
Lagrangian PDF is
\begin{equation}
\label{qvalue}
Q(\eta) = Q_0(f^{-1}(\eta)) {df^{-1}(\eta) \over d \eta},
\end{equation}
where $f^{-1}$ is the inverse of the Lagrangian mapping given in equation
(\ref{mapping}).
Using the relations between the Lagrangian and Eulerian distributions functions,
we can express the final Eulerian density distribution in terms of the initial
Eulerian distribution:
\begin{equation}
\label{pvalue}
P(\eta) = P_0(f^{-1}(\eta)) {f^{-1}(\eta)\over \eta} {df^{-1}(\eta) \over d \eta}.
\end{equation}

All of these equations can be simplified by taking our initial epoch sufficiently
early that $\eta_0 \approx 1$ and $P_0(\eta_0) \approx Q_0(\eta_0)$.
In particular, equation (\ref{pvalue}) becomes
\begin{equation}
\label{p2value}
P(\eta) = P_0(f^{-1}(\eta)) {1\over \eta} {df^{-1}(\eta) \over d \eta}.
\end{equation}

Note that in order for $P$ and $Q$ to represent probability distribution functions,
they must satisfy
\begin{equation}
\label{normalQ}
\int Q(\eta) d\eta = 1,
\end{equation}
and
\begin{equation}
\label{normalP}
\int P(\eta) d\eta  = 1.
\end{equation}
Equation (\ref{normalQ}) is equivalent to mass conservation and is automatically satisfied
by local Lagrangian mappings of the form given in equation (\ref{mapping}).  Equation (\ref{normalP})
gives the conservation of Eulerian volume, and it is not automatically satisfied.
Assuming that our initial distribution functions $P_0$ and $Q_0$ are correctly normalized,
then equation (\ref{normalP}) is satisfied by the Zel'dovich approximation and our
planar approximation; it is not satisfied by equation (\ref{general}) for $\alpha \ne 1$,
and therefore the spherical and exact approximations fail this test.  Hence,
we cannot use equations (\ref{spherical}) or (\ref{1.5}) as written as valid approximations
for the evolution of the density.  To correct this problem,
we modify equation (\ref{mapping}) to read 
\begin{equation}
\eta(\qvec,t) = N(t)f(\eta_0(\qvec),t),
\end{equation}
where $N(t)$ is a time-dependent function given by substituting equation (\ref{p2value}) into our normalization
condition [equation (\ref{normalP})]:
\begin{equation}
\label{renormal}
N(t) = \int {1\over f(\eta_0,t)} P_0(\eta_0) d\eta_0 = \langle {1\over f(\eta_0,t)} \rangle
\end{equation}
With the correct normalization, our general mapping given by eq. (\ref{general})
becomes
\begin{equation}
\label{general2}
\eta(\qvec,t) = {\langle(1-D(t) \delta_0(\qvec)/\alpha)^\alpha \rangle
\over (1-D(t) \delta_0(\qvec)/\alpha)^\alpha}.
\end{equation}
Note that our $N(t)$ resembles the multistreaming factor $N_s$ discussed in
Kofman et al. (1994), although our factor has been introduced as
a mathematical construct in order to
keep our probabilities normalized.

\vskip 1 cm

\noindent{\bf 3 PERTURBATIVE RESULTS}

\noindent{\bf 3.1 Gaussian Initial Conditions}

In order to test our mappings for agreement with the true gravitational
evolution of the PDF, we will first consider the quasilinear case
$|\delta(t)| \la 1$.  In this limit a number of perturbative results
are known for the evolution of Gaussian initial conditions, and a few for
non-Gaussian initial conditions, which can be compared with the perturbative
predictions of our local Lagrangian approximations.

The density field can be described in terms of the cumulants $\kappa_p$,
which are functions of the moments of the density field $\langle \delta^p \rangle$.
The first few cumulants are given
by (see Stuart \& Ord 1987 for a more detailed discussion)
\begin{eqnarray}
\label{cumulants}
\kappa_2 & = & \langle \delta^2 \rangle \nonumber\\
\kappa_3 & = & \langle \delta^3 \rangle  \nonumber\\
\kappa_4 & = & \langle \delta^4 \rangle - 3 \langle \delta^2 \rangle^2
\end{eqnarray}
For Gaussian initial conditions,
it is possible to show
that
the cumulants $\kappa_p$ of the evolved density field
satisfy $\kappa_p/\sigma^{2(p-1)} \rightarrow $ constant in the limit $\sigma \rightarrow 0$,
where $\sigma = \langle \delta^2 \rangle^{1/2}$, and
the constants are called the hierarchical amplitudes, denoted $S_p$:
\begin{equation}
\label{Sp}
S_p = \kappa_p/\sigma^{2(p-1)}
\end{equation}
This result was first derived for the skewness ($\kappa_3$) by Peebles (1980),
for the kurtosis ($\kappa_4$) by Fry (1984), and a method for calculating the
full hierarchy of the $S_p$ was derived by Bernardeau (1992).  Our local
Lagrangian approximations also produce a hierarchical clustering pattern
(i.e., the cumulants of the evolved distribution satisfy equation \ref{Sp}).
Of course, there is little point in using approximations to derive $S_p$ in
the Gaussian case, since these values can be calculated exactly.  However,
it is precisely because the $S_p$'s are known exactly that
the calculation of $S_p$ for the Gaussian case can be used to estimate the
accuracy and general behavior of our approximations.  This calculation also
serves as a warm-up for the case of non-Gaussian initial conditions, for which
no general results for $S_p$ have been derived (although see Fry \& Scherrer 1994;
Chodorowski \& Bouchet 1996).

First note that all of our mappings (equations \ref{planar}, \ref{spherical},
\ref{1.5}) can be expressed in the form 
\begin{equation}
\eta(\qvec,t) = f[\delta_l(\qvec)],
\end{equation}
where $\delta_l$ is the linearly-evolved density:
$\delta_l(\qvec) \equiv D(t) \delta_0(\qvec)$.
Consider first the planar mapping given by
equation (\ref{planar}).  Taking $\eta_0(\qvec) = 1$
we obtain:
\begin{equation}
\label{planarexpansion}
\eta(\qvec,t) = 1 + \delta_l(\qvec) + \delta_l(\qvec)^2 + \delta_l(\qvec)^3 + ...
\end{equation}
To calculate $S_3$, we need to find $\kappa_3 = \langle \delta^3 \rangle_E$,
where we now distinguish between Eulerian averages (effectively taken over volume
and denoted with a subscript $E$) and Lagrangian averages
(taken over mass and denoted with a subscript $L$).  Our task
is to express the Eulerian average of powers of $\delta$ [which
are the numbers which enter into equation (\ref{cumulants})]
in terms of the Lagrangian average of powers of $\eta$ [which can
be derived from equation (\ref{planarexpansion})].
For powers of $\eta$, the relation between the Eulerian and Lagrangian averages
takes the simple form
\begin{equation}
\label{EL}
\langle \eta^n \rangle_E = \langle \eta^{n-1} \rangle_L
\end{equation}
which follows directly from equation (\ref{PQ}).
Then we can express $\kappa_3$ as
\begin{eqnarray}
\label{skewexpansion}
\kappa_3 &=& \langle (\eta-1)^3 \rangle_E \nonumber\\
         &=& \langle \eta^2 \rangle_L - 3 \langle \eta \rangle_L + 2
\end{eqnarray}
Combining equation (\ref{skewexpansion}) with equation (\ref{planarexpansion}),
we obtain, for the planar approximation,
\begin{equation}
\kappa_3 = \sum_{j=1}^\infty \langle (j-2)\delta_l^j \rangle.
\end{equation}
For Gaussian initial conditions,
\begin{eqnarray}
\label{gaussaverage}
\langle \delta_l^j \rangle &=& (j-1)!! \sigma^j ~~(j~{\rm even}), \nonumber\\
     &=& 0 ~~(j~{\rm odd}),
\end{eqnarray}
where $\sigma_l \equiv \langle \delta_l^2 \rangle^{1/2} = D(t) \sigma_0$
is the linearly-evolved rms fluctuation.
Then we end up with
\begin{equation}
\kappa_3 = \sum_{n=2}^\infty 2(n-1)(2n-1)!! \sigma_l^{2n},
\end{equation}
so that
\begin{equation}
S_3(\sigma_l) = \sum_{n=2}^\infty 2(n-1)(2n-1)!! \sigma_l^{2n-4} = 6 + 60 \sigma_l^2 + O(\sigma_l^4),
\end{equation}
and $S_3(0) = 6$.
This result has been derived previously by Bernardeau \& Kofman (1995) using more complex
techniques, but our $\sigma^2$ term differs from theirs.  At this point, we must be
careful about our definition of the hierarchical amplitudes.  In equation (\ref{Sp}),
we have used the {\it linearly-evolved} rms fluctuation $\sigma_l$, rather than the true
rms fluctuation $\sigma = \langle \delta^2 \rangle^{1/2}$.  If
instead we use the true rms fluctuation, we can define a new set of hierarchical
amplitudes, given by:
\begin{equation}
\label{trueSp}
\widetilde S_p(\sigma) = \kappa_p / \langle \delta^2 \rangle^{p-1}
\end{equation}
This distinction is usually ignored, because $S_p(0) = \widetilde S_p(0)$ for Gaussian
initial conditions.  However, when expanding $S_p$ to higher order, or when
dealing with non-Gaussian initial conditions (Fry \& Scherrer 1994), the distinction must be made.  Note that $\widetilde S_p$ is closer to what observers actually measure.
When we expand
$\langle \delta^2 \rangle_E =  \langle \eta \rangle_L - 1$, we obtain
\begin{equation}
\sigma^2 = \sigma_l^2 + 3 \sigma_l^4 + O(\sigma_l^6).
\end{equation}
Substituting this into equation (\ref{trueSp}) and reexpressing everything in terms
of $\sigma$ rather than $\sigma_l$, we obtain
an expression
for $\widetilde S_p$ which agrees with Bernardeau \& Kofman:
\begin{equation}
\label{skewtilde}
\widetilde S_3(\sigma) = 6 + 24 \sigma^2 + O(\sigma^4)
\end{equation}
Note that because $\sigma^2 = \sigma_l^2 + O(\sigma_l^4)$, it does not matter if we use
$\sigma_l$ or $\sigma$ on the right-hand side of equation (\ref{skewtilde}), although
it would make a difference if we expanded out to fourth order in $\sigma$.

We can similarly expand the expression for $\kappa_4$ to obtain
\begin{equation}
\label{kurtexpansion}
\kappa_4 = \langle \eta^3 \rangle_L - 4 \langle \eta^2 \rangle_L - 3 \langle \eta \rangle_L^2
+ 12 \langle \eta \rangle_L - 6 
\end{equation}
Again, substituting equation (\ref{planarexpansion}) for $\eta$ in equation
(\ref{kurtexpansion}), expanding out term by
term, and taking the averages appropriate to the Gaussian initial conditions from equation
(\ref{gaussaverage}), we obtain
$S_4(0) = 72$,
in agreement with Bernardeau \& Kofman (1995).
%In fact, it is possible to derive
%a general expression for $\langle \delta^p \rangle$ for the planar expansion.  In Appendix A, %we show that
%\begin{equation}
%\langle \delta^p \rangle_E = \sum_{k=0}^\infty \sum_{j=[p/2]???}^\infty
%2^{1-3k} {(p-1)!(2k)! \over (2s-p)! (s-1)! k!} F(-k,s,1/2,1) \sigma_l^{2s},
%\end{equation}
%where $F$ is a hypergeometric function (actually, a Jacobi polynomial, since the
%negative first argument gives a finite series).

Now consider the spherical approximation (equation \ref{spherical}).  This approximation
is complicated by the fact that for Gaussian initial conditions,
$\langle 1/f(\eta_0,t) \rangle \ne 1$, so we have to include the normalizing factor given
by equation (\ref{renormal}).  For Gaussian initial conditions, the normalizing factor in equation
(\ref{renormal}) for the spherical approximation is
\begin{equation}
N = \langle (1-D(t) \delta_0/3)^3 \rangle = 1 + \sigma_l^2/3
\end{equation}
and the spherical mapping becomes:
\begin{equation}
\eta(\qvec,t) = {1+\sigma_l^2/3\over (1-\delta_l/3)^3}.
\end{equation}
Substituting this mapping into equations (\ref{skewexpansion}) and
(\ref{kurtexpansion}) and taking the
appropriate averages for the Gaussian initial distribution from equation
(\ref{gaussaverage}) we obtain
\begin{eqnarray}
S_3(0) & = & 4 \\
S_4(0) & = & 272/9
\end{eqnarray}
Oddly, these are identical to the hierarchical amplitudes obtained for the
full Zel'dovich approximation (see, for example, Munshi et al. 1994; Bernardeau \& Kofman 1995). In fact, we can show that
the spherical approximation and the Zeldovich approximation have
identical values for {\it all} of the hierarchical amplitudes $S_p(0)$.
To do this, we introduce the vertex generating function for
the density field (Bernardeau 1992)
\begin{equation}
\label{vertex1}
G_\delta(\tau) = \sum_{n=1}^\infty {\nu_n \over n!} \tau^n,
\end{equation}
where
\begin{equation}
\label{vertex2}
\nu_n = {\int \langle \delta^{(n)}(\xvec) \delta^{(1)}(\xvec_1)...\delta^{(1)}(\xvec_n)\rangle_c
d^3\xvec ~ d^3 \xvec_1...d^3\xvec_n
\over(\int \langle \delta^{(1)}(\xvec)\delta^{(1)}(\xvec^\prime) \rangle
d^3\xvec ~ d^3 \xvec^\prime)^n},
\end{equation}
and $\delta^{(n)}$ is the $n_{th}$-order expansion of $\delta$.
[Our convention for the sign of $\tau$ is the same as that of Munshi et al. (1994)
and the opposite of Bernardeau (1992) and Bernardeau \& Kofman (1995)].
The values
of the $\nu_n$ totally determine $S_p(0)$, through the relations:
\begin{eqnarray}
\label{Snu}
S_3(0) &=& 3 \nu_2, \nonumber \\
S_4(0) &=& 4 \nu_3 + 12 \nu_2^2,
\end{eqnarray}
and so on (Bernardeau 1992).

In the Appendix, we demonstrate an extremely useful result for local Lagrangian mappings
with Gaussian initial conditions:
for a given mapping $\eta(\qvec,t) = N(t) f[\delta_l(\qvec)]$, where $\delta_l$ is
the linearly-evolved density, the vertex generation function is just given by the
same Lagrangian
mapping without the normalizing function:
\begin{equation}
\label{Gdelta}
G_\delta(\tau) = f(\tau) - 1
\end{equation}
[In fact, the same result holds for local Eulerian mappings, but these are not the subject
of this paper].  For the spherical approximation, equation (\ref{Gdelta}) gives
\begin{equation}
G_{\delta}(\tau) = {1\over (1-\tau/3)^3} - 1
\end{equation}
which is identical to $G_\delta(\tau)$ for the Zeldovich approximation
(Munshi, et al.). Thus, $S_p(0)$ for our spherical approximation and $S_p(0)$ for
the Zeldovich approximation will be identical for all $p$.  This does not mean
that the spherical approximation and the Zeldovich approximation produce identical
evolved PDF's.
The reason is that although $S_p(0)$ is the same for these
two approximations, the $S_p(\sigma)^\prime s$ differ at higher order in $\sigma$.  The importance
of considering higher-order $\sigma$ terms has been emphasized by
Bernardeau \& Kofman (1995), and Scoccimarro \& Frieman (1996) have recently calculated
the higher-order corrections for the Zeldovich approximation.  Scoccimarro \&
Frieman obtain, for $S_3(\sigma)$:
\begin{eqnarray}
S_3(\sigma_l) &=& 4 + (1112/75) \sigma_l^2 + O(\sigma_l^4) \\
\widetilde S_3(\sigma) &=& 4 + (352/75) \sigma^2 + O(\sigma^4)
\end{eqnarray}
In comparison, our spherical approximation gives:
\begin{eqnarray}
S_3(\sigma_l) &=& 4 + (286/27) \sigma_l^2 + O(\sigma_l^4) \\
\widetilde S_3(\sigma) &=& 4 + (118/27) \sigma^2 + O(\sigma^4).
\end{eqnarray}
Although the higher-order contributions to $S_3$ are of similar magnitude in the two cases,
they are not identically equal, so the spherical approximation and Zeldovich approximation
give different density distributions.  This result demonstrates the
importance of higher-order calculations:
two density fields can have identical $S_p(0)$ for all $p$ and yet have different PDF's.

The result given in equation (\ref{Gdelta}) leads naturally to our ``exact" approximation
(equation \ref{1.5}).
An expression for $G_\delta(\tau)$ can be calculated in parametric form for the
case of the exact evolution
of the density field (Bernardeau 1992; Munshi et al. 1994; Bernardeau \& Kofman 1995):
\begin{eqnarray}
\label{parametric1}
G_\delta(\tau) &=& {9 \over 2}{(\theta-\sin\theta)^2 \over
(1-\cos\theta)^3} - 1, \nonumber\\
\tau &=& {3\over 5} [{3\over 4}(\theta-\sin\theta)]^{2/3},
\end{eqnarray}
for $\tau > 0$, and
\begin{eqnarray}
\label{parametric2}
G_\delta(\tau) &=& {9 \over 2}{(\sinh\theta-\theta)^2 \over
(\cosh\theta-1)^3} - 1, \nonumber\\
\tau &=& - {3\over 5} [{3\over 4}(\sinh\theta-\theta)]^{2/3},
\end{eqnarray}
for $\tau < 0$.
Bernardeau (1992) and Bernardeau \& Kofman (1995) have noted that an excellent
approximation for $G_\delta(\tau)$ for the case of exact evolution is given by
\begin{equation}
G_\delta(\tau) = (1 - 2\tau/3)^{-3/2} - 1.
\end{equation}
Now we can argue backwards from our result in equation (\ref{Gdelta}):  the local
Lagrangian mapping
\begin{equation}
\label{exact}
\eta(\qvec,t) = {\langle(1-2\delta_l/3)^{3/2}\rangle
\over (1- 2\delta_l/3)^{3/2}}
\end{equation}
will produce a hierarchy of $S_p(0)$ very close to the results of exact evolution.
Hence, the local Lagrangian mapping given in equation (\ref{exact}) should
provide a good approximation to the true evolution in the quasi-linear regime.
As expected, the lowest order hierarchical amplitudes for this approximation
are in excellent agreement with the exact values:
our approximation gives $S_3(0) = 5$, $S_4(0) = 440/9 \approx 48.9$, compared to
$S_3(0) = 4.9$ and $S_4(0) = 45.9$ for the exact quasi-linear evolution.

We note in passing that using equations (\ref{parametric1})-(\ref{parametric2}),
along with equation (\ref{Gdelta}), it is possible to derive a local
Lagrangian approximation in parametric form which {\it exactly} reproduces the hierarchical amplitudes
$S_p(0)$ for all $p$.  However, the extra complexity involved in the
parametric representation, plus the fact
that higher order terms will diverge from their correct values anyway, probably makes
this approximation less useful than our ``exact" approximation.

Now let us evaluate the usefulness of the approximations which we have derived.
A comparison of $G_\delta(\tau)$ for the planar and spherical approximations
with $G_\delta(\tau)$ for the case of exact evolution, expanded out in a power
series (Munshi et al. 1994) shows that the coefficients in the expansion for
the planar case are larger than the corresponding coefficients for exact evolution,
while the opposite is true for the spherical approximation.  Thus, $S_p(0)$
for the planar approximation gives an upper bound on the true $S_p(0)$, while
$S_p(0)$ for the spherical approximation gives a lower bound.  In that sense,
these two approximations bound the true evolution of the PDF:  the planar
approximation gives a PDF which deviates more strongly from a Gaussian than
the true evolution, while the spherical approximation yields a PDF which
deviates less from a Gaussian.  This makes physical sense, since these two
approximations correspond to planar collapse and spherical collapse, respectively,
and the true evolution should lie somewhere in between.  The ``exact" approximation,
on the other hand, should provide a reasonable approximation to the evolution
of the PDF for quasi-linear evolution.

All of these calculations are valid only for Gaussian initial conditions.  However,
it is plausible that our three approximations can be extended to provide
some insight into the evolution of the PDF for the case of non-Gaussian initial
conditions:  arguing in analogy with the Gaussian case,
we expect the planar and spherical approximations to provide upper and lower
bounds on the deviation of the PDF from the initial conditions, while the
``exact" approximation should give a good estimate of the overall evolution.
More importantly, any features in the evolution shared by all three approximations
are likely to be true characteristics of the gravitational evolution of the PDF.

\medskip

\noindent{\bf 3.2 Effects of Smoothing}

\medskip

Before venturing into the murky world of non-Gaussian initial conditions,
we must consider the effects of smoothing.  Our approximations and results in
the previous section apply only to the unsmoothed density field, while
it is the smoothed density field which is actually observed.  Since our local
Lagrangian approximations give only the PDF and
do not provide a prescription for actually moving
the matter around, there is in principle no way to derive smoothed versions of
them.  However, we can derive plausible ``smoothed" approximations which
give the same hierarchical amplitudes as the smoothed mappings would.

Our argument is based on the calculations of Bernardeau (1994), who showed
that for a spherical tophat window function,
there is a simple relation between $G_\delta(\tau)$ for a particular density field,
and $G^S_\delta(\tau)$ for the corresponding smoothed
density field.  For simplicity, we consider only a density field
with a power-law power spectrum
$P(k) \propto k^n$, for which Bernardeau (1994) obtained the implicit equation
\begin{equation}
\label{smooth}
G^S_\delta(\tau) = G_\delta(\tau[1+G^S_\delta(\tau)]^{-(n+3)/6}).
\end{equation}
Consider the vertex function of the density field produced by first applying
a given local Lagrangian mapping, and then smoothing.  Using equation (\ref{smooth}),
we can generate a ``smoothed" local Lagrangian mapping which produces a final density 
field with the same vertex generating function.
To do this, we simply use equation (\ref{Gdelta}):
for local Lagrangian approximations, a transformation of the vertex generating function
corresponds to the same transformation of the Lagrangian mapping.
Thus, for a given mapping $\eta(\qvec,t) = N(t) f(\delta_l)$, the ``smoothed" mapping
is given by
\begin{equation}
\label{Lagsmooth}
\eta_S =  f^S(\delta_l) = f(\delta_l[f_S(\delta_l)]^{-(n+3)/6})
\end{equation}
where $f^S(\delta_l)$ must then be multiplied by the normalizing factor specified by equation
(\ref{normalP}).
Note that this does
{\it not} mean that $f^S(\delta_l)$ corresponds
to the density field derived by applying the mapping $\eta = f(\delta_l)$ and then smoothing;
rather, $f_S(\delta_l)$ will produce a density field with the same
hierarchical amplitudes $S_p(0)$ as the field derived by first applying $f(\delta_l)$ and then
smoothing.  Hence, we expect $f^S(\delta_l)$ given by equation (\ref{Lagsmooth}) to provide a good approximation
to the PDF of the smoothed density field.

We will confine our attention to the case $n=-1$, because it corresponds
roughly to the the slope of the CDM power spectrum on galaxy clustering
scales, and this choice allows us to compare with previous work.
For
the spherical and ``exact" approximations, this case produces
particularly simple smoothed mappings.
For the spherical mapping, equation (\ref{Lagsmooth}) applied to the mapping
given by equation (\ref{spherical}) gives (Bernardeau \& Kofman 1995)
\begin{equation}
\eta_S = (1+\delta_l/3)^3 \langle (1 + \delta_l/3)^{-3} \rangle
\end{equation}
while for the exact approximation, equations (\ref{Lagsmooth}) and (\ref{1.5})
yield (Bernardeau 1994):
\begin{equation}
\eta_S = (\delta_l/3 + \sqrt{1+\delta_l^2/9})^{3} \langle (\delta_l/3 + \sqrt{1+\delta_l^2/9})^{-3} \rangle 
\end{equation}
For the planar approximation, we obtain a cubic equation for $\eta_S$,
which yields
\begin{equation}
\eta_S = [g^{1/3} + \delta_l^2 /9g^{1/3} + \delta_l/3]^3 \langle
[g^{1/3} + \delta_l^2 /9g^{1/3} + \delta_l/3]^{-3} \rangle ,
\end{equation}
where $g(\delta_l)$ is defined by
\begin{equation}
g = 1/2 + \delta_l^3/27 + (12 \delta_l^3 + 81)^{1/2} /18.
\end{equation}
We can now apply both the smoothed and unsmoothed approximations
to the case of non-Gaussian initial conditions.

\medskip

\noindent{\bf 3.3 Non-Gaussian Initial Conditions}

We now repeat our perturbative calculations from Section 3.1 for
the case of non-Gaussian initial conditions.  The methods are identical
to those used in Section 3.1; the only difference is that
for non-Gaussian initial conditions, all terms of the form
$\langle \delta_l^p \rangle$ must be retained, rather than
being reduced to various powers of $\sigma$ as we did for Gaussian
initial conditions (equation \ref{gaussaverage}).  This alters
not just our final expression derived from our mappings,
but also changes the normalization factor for each mapping.

Consider
first the rms fluctuation $\delt2$.  The most
interesting question one can ask about $\delt2$
is whether the first correction gives growth which is faster or slower
than linear.
For
Gaussian initial conditions, the first correction to linear theory
is of order $\sigma^4$.  This correction has recently been examined
in detail by Lokas et al. (1996) and Scoccimarro \& Frieman (1996); they find that for unsmoothed
density fields,
\begin{equation}
\langle \delta^2 \rangle = \sigma_l^2 + 1.8 \sigma_l^4
\end{equation}
For smoothed initial conditions, the sign of the $\sigma^4$ term
depends on the initial power index; it is negative for $n \ge -1$
but positive for $n=-2$.

As noted by Fry \& Scherrer (1994), the first correction to $\delt2$
for non-Gaussian initial conditions is of order $\sigma^3$, rather
than $\sigma^4$, suggesting that non-Gaussian initial conditions
should give an earlier divergence from linear behavior.  Fry \& Scherrer (1994)
found that for arbitrary initial conditions,
\begin{equation}
\label{delta2}
\delt2 = \sigma_l^2 + {13 \over 21} \langle \delta_l^3 \rangle + {4 \over 7}
I[\zeta_0],
\end{equation}
where $I[\zeta_0]$ is an integral over the initial three-point function
of the non-Gaussian distribution.  If the second term in equation
(\ref{delta2}) dominates the third term or has the same sign as the third
term, then the sign of the initial skewness determines whether
the rms density evolves more or less rapidly than linear, with positive
skewness models evolving more rapidly and negative skewness models
less rapidly.  However, all of these results apply only to the unsmoothed
density field.

Using our approximations from Section 2 and our perturbative methods
from Section 3, we obtain, for the unsmoothed density field,
the following expressions for $\delt2$:
\begin{eqnarray}
\delt2 &=& \sigma_l^2 + \langle \delta_l^3 \rangle
+ \langle \delta_l^4 \rangle + O(\sigma^5), ~~(\rm{planar}),\\
\delt2 &=& \sigma_l^2 + {2\over 3}\langle \delta_l^3 \rangle
+ {53\over 108}\langle \delta_l^4 \rangle + {5\over 36} \sigma_l^4
+ O(\sigma^5), ~~(\rm{exact}),\\
\delt2 &=& \sigma_l^2 + {1\over 3}\langle \delta_l^3 \rangle
+ {5\over 27}\langle \delta_l^4 \rangle + {2\over 9} \sigma_l^4
+ O(\sigma^5), ~~(\rm{spherical}).
\end{eqnarray}
For the Gaussian case, we see that the first correction term is
$3 \sigma^4$ for the planar case, $1.61\sigma^4$ for the exact
approximation, and $0.78 \sigma^4$ for the spherical approximation.
Once more, we see that the planar and spherical cases bracket the
exact perturbative result, while the exact approximation comes
close to the exact perturbative value.  For the non-Gaussian case,
the expression for the exact evolution is non-local, as shown
in equation (\ref{delta2}).  However, in the limit where the long-range
correlations in the initial density field are small,
equation
(\ref{delta2}) reduces to $\delt2 = \langle \delta_l^2 \rangle + (13/21) \langle \delta_l^3 \rangle$.
The nonlinear correction term in this case is nearly identical to the correction
term in the exact approximation, $(2/3) \langle \delta_l^3 \rangle$,
and is again bracketed by the correction terms for the
planar and spherical approximations.
In all three of our approximations, the lowest-order nonlinear correction
is a positive multiple of the initial skewness, so that positive-skewness models
have larger rms densities and negative-skewness models have smaller rms
densities than predicted by linear theory, in agreement with the conjecture
of Fry \& Scherrer (1994).

The simulations of Weinberg and Cole (1992) do not unambiguously support this conclusion for the
case of smoothed density fields, so we consider what happens when we use our
smoothed mappings from the previous section.  Using the mappings for $n=-1$ derived in the
previous section, we find:
\begin{eqnarray}
\delt2 &=& \sigma_l^2 + {1\over3}\langle \delta_l^3 \rangle
+ {1\over 9}\langle \delta_l^4 \rangle + {2\over 9}\sigma_l^4
+O(\sigma^5), ~~(\rm{planar}),\\
\delt2 &=& \sigma_l^2
+ {5\over 108}\langle \delta_l^4 \rangle + {1\over 4} \sigma_l^4
+ O(\sigma^5), ~~(\rm{exact}),\\
\delt2 &=&  \sigma_l^2 - {1\over 3}\langle \delta_l^3 \rangle
+ {5\over 27}\langle \delta_l^4 \rangle + {2\over 9} \sigma_l^4
+ O(\sigma^5), ~~(\rm{spherical}).
\end{eqnarray}
We see that the effect of smoothing for $n=-1$ is to significantly reduce
the growth of $\delt2$ as compared with the unsmoothed case.  In particular,
we can draw no unambiguous conclusions regarding the dependence of $\delt2$ on
the sign of the skewness; our three approximations give different answers,
and in our exact approximation, there is no skewness dependence at all.

Now consider the evolution of the skewness itself.  As noted in section 3.1,
we must distinguish between
$S_3(\sigma) = \kappa_3 / \langle \delta_l^2 \rangle^4$
and $\widetilde S_3(\sigma) = \kappa_3 / \langle \delta^2 \rangle^4$.
For non-Gaussian initial conditions these two quantities are not
equal even in the limit $\sigma \rightarrow 0$, because of the terms
containing $\langle \delta_l^3 \rangle $ in the expansion for $\delt2$.
Here we will consider only $\tilde S_3(0)$, because this is the quantity
measured by observers; it is also the definition of skewness used
by Fry \& Scherrer (1994).  Again, using the mappings in Sections
2 and 3.2 along with the expression for $\tilde S_3$ in section 3.1,
we can obtain expressions for $S_3(0)$ for our various approximations.
For comparison with the Gaussian case (and with previous work) it
is convenient to express these results in terms of the linearly-evolved
kurtosis, defined by $\kappa_{4l} \equiv = \langle \delta_l^4 \rangle - 3 \sigma_l^4$,
which vanishes for Gaussian initial conditions.  For the unsmoothed case, we obtain:
\begin{eqnarray}
\widetilde S_3(0) &=& {\langle \delta_l^3 \rangle \over \sigma_l^4} + 6
+ 2 {\kappa_{4l} \over \sigma_l^4}
- 2 {\langle \delta_l^3 \rangle^2 \over \sigma_l^6}, ~~({\rm planar}) \\
\widetilde S_3(0) &=& {\langle \delta_l^3 \rangle \over \sigma_l^4} + 5
+ {3\over 2} {\kappa_{4l} \over \sigma_l^4}
- {4 \over 3} {\langle \delta_l^3 \rangle^2 \over \sigma_l^6}, ~~({\rm exact}) \\
\widetilde S_3(0) &=& {\langle \delta_l^3 \rangle \over \sigma_l^4} + 4
+ {\kappa_{4l} \over \sigma_l^4}
- {2 \over 3} {\langle \delta_l^3 \rangle^2 \over \sigma_l^6}, ~~({\rm spherical}).
\end{eqnarray}
The exact perturbative result for $\widetilde S_3(0)$ for non-Gaussian
initial conditions (Fry \& Scherrer 1994) contains non-local terms involving
integrals over the the initial three- and four-point functions.  However, in
the limit where these terms are small (e.g., for the case of weak correlations
in the initial density field), the exact perturbative result is:
\begin{equation}
\widetilde S_3(0) = {\langle \delta_l^3 \rangle \over \sigma_l^4} + {34\over 7}
+ {10\over 7} {\kappa_{4l} \over \sigma_l^4}
- {26 \over 21} {\langle \delta_l^3 \rangle^2 \over \sigma_l^6}, ~~({\rm exact}).
\end{equation}
This expression is remarkably close to our result for the ``exact approximation",
and it is bracketed by the results for the planar and spherical approximations.

For the smoothed case, we obtain the results:
\begin{eqnarray}
\widetilde S_3(0) &=& {\langle \delta_l^3 \rangle \over \sigma_l^4} + 4
+  {\kappa_{4l} \over \sigma_l^4}
- {2\over 3} {\langle \delta_l^3 \rangle^2 \over \sigma_l^6}, ~~({\rm planar}) \\
\widetilde S_3(0) &=& {\langle \delta_l^3 \rangle \over \sigma_l^4} + 3
+ {1\over 2} {\kappa_{4l} \over \sigma_l^4}, ~~({\rm exact}) \\
\widetilde S_3(0) &=& {\langle \delta_l^3 \rangle \over \sigma_l^4} + 2
+ {2 \over 3} {\langle \delta_l^3 \rangle^2 \over \sigma_l^6}, ~~({\rm spherical}).
\end{eqnarray}
There are no exact perturbative results with which we can compare our
smoothed expressions for $\widetilde S_3$, but they display qualitatively
the correct behavior; the effect of smoothing is to reduce the evolved skewness.

The results of this section for non-Gaussian initial conditions confirm the conclusions
we reached in Section 3.1.  Our ``exact" approximation for the unsmoothed case
agrees well with perturbative results for non-Gaussian initial conditions, while
the planar and spherical results bracket the known perturbative results.
This suggests that our ``exact" approximation should give reasonable results
when we go on to calculate the full evolved PDF for a variety of non-Gaussian initial
conditions, while the planar and spherical approximations may serve as useful
bounds on the evolution.  The situation is murkier with regard to our
smoothed approximations, but they show at least qualitatively the correct behavior.

\vskip 1 cm

\noindent{\bf 4 THE EVOLVED PDF}

We can now apply our local Lagrangian mappings to derive the evolved PDF for any
set of initial conditions using equation (\ref{p2value}).  Since we no
longer assume $\delta_l \ll 1$, we have to modify
our generic mapping given in eq. (\ref{general2}).  Following Kofman et al. (1994)
we take
\begin{equation}
\label{absolute}
\eta(\qvec,t) = {N(t) \over |1-\delta_l/\alpha|^\alpha}
\end{equation}
where the normalization factor $N(t)$ is now given by
\begin{equation}
N(t) = \langle |1-\delta_l/\alpha|^\alpha \rangle
\end{equation}
and the average is taken over the distribution of $\delta_l$.

Consider first the case of Gaussian initial conditions.  Then
for the unsmoothed case, equation (\ref{p2value}) gives
\begin{equation}
\label{Pnosmooth}
P(\eta)d\eta = {1\over \sqrt{2\pi}\sigma_l} \big({\eta\over N}\big)^{{-1/\alpha}-2}
\Big[e^{-\alpha^2[1-(\eta/N)^{-1/\alpha}]^2/2\sigma_l^2} + e^{-\alpha^2[1+(\eta/N)^{-1/\alpha})]^2
/2\sigma_l^2}\Big] {d\eta\over N^2},
\end{equation}
where $N$ is given by
\begin{equation}
N = \int_{-\infty}^\infty {1\over \sqrt{2\pi}\sigma_l}
e^{-\delta_l^2/2\sigma_l^2}|1 - {\delta_l\over \alpha}|^\alpha d\delta_l.
\end{equation}
Note that this expression for $P(\eta)$ bears some resemblence
to the distribution function derived by Bernardeau (1994), [eq. (19)],
using more sophisticated techniques, but the two distributions are different.

For our smoothed mappings, the results are even simpler.  If we assume a mapping
of the form in equation (\ref{absolute}) smoothed with a spherical tophat
window function, with a $k^n$ power spectrum, we obtain
from equation (\ref{Lagsmooth}) the following relation
between the evolved value for $\eta$ and the linear perturbation $\delta_l$:
\begin{equation}
\label{smoothmap}
\delta_l = \alpha\Big[({\eta\over N})^{(n+3)/6} - ({\eta\over N})^{-1/\alpha + (n+3)/6}\Big],
\end{equation}
which can be substituted into the Gaussian expression for $\delta_l$, with the appropriate
normalizing factor, to obtain $P(\eta)$.  For example,
for $n=-1$ we obtain
\begin{equation}
\label{Pgeneral}
P(\eta) d\eta = {1\over \sqrt{2\pi}\sigma_l} \alpha[({1\over 3} ({\eta\over N})^{-5/3}
+({1\over \alpha} - {1\over 3})({\eta\over N})^{-{1/\alpha} - {5/3}}]
\exp\Big(-\alpha^2[({\eta\over N})^{1/3}-({\eta\over N})^{1/3-1/\alpha}]^2/2\sigma_l^2\Big)
{d\eta\over N^2}
\end{equation}
For the ``exact" approximation ($\alpha = 3/2$), this reduces to
\begin{equation}
\label{Pexact}
P(\eta) d\eta = {1\over 2 \sqrt{2\pi}\sigma_l} [({\eta\over N})^{-5/3}
+({\eta\over N})^{-7/3}]
\exp\Big(-{9\over 8 \sigma_l^2}[({\eta\over N})^{1/3}-({\eta\over N})^{-1/3}]^2\Big)
{d\eta\over N^2},
\end{equation}
where $N$ in this case is given by
\begin{equation}
\label{Nexact}
N = \int_{-\infty}^\infty {1\over \sqrt{2\pi}\sigma_l}
e^{-\delta_l^2/2\sigma_l^2} \Bigg[{1\over 3} \delta_l+
\sqrt{1+{\delta_l^2 \over 9}}\Bigg]^{-3} d\delta_l
\end{equation}

We can compare these results with numerical simulations of gravitational clustering.
We have
used the simulations of Weinberg (Juszkiewicz et al. 1995;
Lokas et al. 1995) with power-law initial conditions,
smoothed with a spherical tophat.  In Figure 1, we show the case $n=-1$, for $\sigma_l =$
0.5, 0.75, and 1.  The points are the results of averaging eight numerical
simulations with 1-$\sigma$ error bars, and the solid curve is the smoothed ``exact"
approximation, equation (\ref{Pexact}).  For $\sigma_l = 0.5$, the agreement is remarkable, particularly
given the simplicity of the approximation which led to equation (\ref{Pexact}).
There is still reasonable
agreement at $\sigma_l = 0.75$, but the approximation begins
to break down at this point, and agreement is poor for $\sigma_l = 1.0$.
These results are not surprising, since the ``exact" approximation
was designed to mimic the exact hierarchical amplitudes in the
limit where $\sigma_l \rightarrow 0$, and it can be expected
to break down when the contributions to $S_p$ of higher order in $\sigma_l$
become important.  In fact, Scoccimarro \& Frieman (1996) have argued
that this should occur at $\sigma_l^2 \sim 1/2$, which is precisely when our
approximation appears to break down in Fig. 1.  In Fig. 2, we
show the planar and spherical approximations along with the numerical
simulations for $\sigma_l = 0.5$ and $\sigma_l = 0.75$.  For $\sigma_l = 0.5$,
the planar and spherical approximations do indeed bound the numerical
results, but the spherical approximation breaks down completely
at $\sigma_l = 0.75$.  In fact, it is easy to show from equation (\ref{Pgeneral})
that the spherical approximation breaks down (in the sense of
no longer having central maximum near $\eta = 1$) at
$\sigma_l^2 = 9/20$.  Again, this is near the point
at which Scoccimarro and Frieman (1996) predict a breakdown of perturbation
theory.

The importance of using our smoothing method derived
from Bernardeau (1994) is illustrated in Fig. 3.  There we show,
for $\sigma_l = 0.5$, all three of our approximations for the unsmoothed
PDF.   All three of them differ strongly from the numerical results,
showing greater deviation from the original Gaussian.
This is not surprising, since these approximations are only appropriate
for the PDF measured for an unsmoothed density field, or for
a smoothed density field with initial
power spectrum $P(k) \propto k^{-3}$.

Finally, we note that our results in the Appendix indicate that there
is also a local Eulerian mapping which has a vertex generating function
equal to $G_\delta(\tau) = 1/(1- 2\tau /3)^{3/2}$ and which should
therefore also give an excellent approximation to the evolved PDF
in the quasilinear regime.  This Eulerian approximation
is given by taking $\eta(\xvec,t) = f[\delta_l(\xvec)]$, with the mapping
given by equation (\ref{absolute}) or (\ref{smoothmap}) for the unsmoothed and smoothed
density fields, respectively.  The final result is a PDF which
looks the same as equation (\ref{Pnosmooth}) or equation (\ref{Pgeneral}), but missing the factor of $1/\eta$
which transforms the Lagrangian PDF $Q(\eta)$ into the Eulerian PDF $P(\eta)$.
So, for example, the final $P(\eta)$ for the Eulerian
version of the ``exact" approximation smoothed with a tophat window function for $n=-1$
is simply obtained by
multiplying equation (\ref{Pexact}) by $\eta$:
\begin{equation}
\label{Plagexact}
P(\eta) d\eta = {1\over 2 \sqrt{2\pi}\sigma_l} [({\eta\over N})^{-2/3}
+({\eta\over N})^{-4/3}]
\exp\Big(-{9\over 8 \sigma_l^2}[({\eta\over N})^{1/3}-({\eta\over N})^{-1/3}]^2\Big)
{d\eta\over N}
\end{equation}
where the normalization factor is now given by:
\begin{equation}
N = \int_{-\infty}^\infty {1\over \sqrt{2\pi}\sigma_l}
e^{-\delta_l^2/2\sigma_l^2} \Bigg[{1\over 3} \delta_l+
\sqrt{1+{\delta_l^2 \over 9}}\Bigg]^{3} d\delta_l
\end{equation}
rather than the expression in equation (\ref{Nexact}).
This Eulerian approximation is compared with the numerical results for $\sigma_l = 0.5$
in Figure 4.  The agreement between this approximation and the numerical
simulations is excellent.  It may seem implausible that both the Eulerian
and Lagrangian local mappings could produce nearly the same final $P(\eta)$,
since the distributions given by equations (\ref{Pexact}) and (\ref{Plagexact})
differ by a factor of $\eta$.  However, this difference is compensated
by the different values for $N$ used in the two equations.  In fact,
our argument in the appendix indicates that these two approximations
should give equally good agreement with the evolved PDF in the limit
where $\sigma_l \ll 1$.

Now consider the evolution of non-Gaussian initial conditions.  Our approximation
can be applied in an elementary way to any initial density distribution,
but there are an infinite set of distributions to choose from.
To explore the differences which the initial skewness and kurtosis
make in the evolution, we have chosen four representative distributions;
one each with positive and negative skewness, and two symmetric (zero skewness)
distributions with positive and negative kurtosis.  The distributions
we examine below represent some extreme cases and are not physically
motivated.  However, they give a general idea of the effects of positive
and negative skewness or kurtosis on the evolution of the PDF.

We will consider only the smoothed exact approximation with $\sigma_l \le 0.75$,
which we know produces results in good agreement with the true PDF
for the Gaussian case, and for definiteness we will take $n=-1$.
For this case, the mapping in equation (\ref{smoothmap}) becomes
\begin{equation}
\delta_l = {3 \over 2} \Big[({\eta\over N})^{1/3} - ({\eta\over N})^{-1/3}\Big],
\end{equation}
with $N$ given by:
\begin{equation}
N = \int_{-\infty}^\infty P(\delta_l) \Bigg[{1\over 3} \delta_l+
\sqrt{1+{\delta_l^2 \over 9}}\Bigg]^{-3} d\delta_l,
\end{equation}
where $P(\delta_l)$ is the initial (non-Gaussian) distribution
which we are evolving.

For the case of positive skewness, a simple choice is the gamma
distribution with zero mean and variance $\sigma_l^2$:
\begin{equation}
\label{gamma}
P(\delta_l)d\delta_l = {\nu^{\nu/2} \over \Gamma(\nu) \sigma_l} ({\delta_l \over \sigma_l} +
\sqrt{\nu})^{\nu-1} e^{-\nu -\sqrt{\nu} \delta_l/\sigma_l} d \delta_l,
\end{equation}
where each value of $\nu$ defines a different gamma distribution.
A mirror image negative skewness distribution with zero mean can be obtained by simply changing $\delta_l$ to $-\delta_l$ in equation (\ref{gamma}).
For definiteness, we take $\nu = 3$, and our results for positive
and negative initial skewness are shown in figures 5 and 6, at
$\sigma_l = 0.2$, 0.5, and 0.75.  Both density distributions
show qualitatively the expected evolution, i.e., the development of
increasing skewness and a large positive tail.  For the positive
skewness case (figure 5) this does not represent a major change
in the shape of the distribution function.  The effect is
more dramatic in figure 6, where the distribution function
with negative initial skewness flips into
a PDF with positive skewness, as expected.
Oddly, the function with negative initial skewness develops
a larger tail at large $\eta$ than does the function with positive
initial skewness.

For symmetric distribution functions we have chosen two ``extreme" representative
models, the bilateral exponential:
\begin{equation}
P(\delta_l)d\delta_l = {1 \over \sqrt{2} \sigma_l} e^{-\sqrt{2} |\delta_l| / \sigma_l} d\delta_l,
\end{equation}
which has large positive kurtosis, and
the uniform distribution
\begin{eqnarray}
P(\delta_l)d\delta_l &=& {1 \over 2 \sqrt{3} \sigma_l} d\delta_l,~~~|\delta_l| < \sqrt{3} \sigma_l, \\
       &=& 0~~~~~~~~~({\rm otherwise}),
\end{eqnarray}
with large negative kurtosis.
Neither of these can be considered a realistic initial distribution, but
they illustrate the effect of large kurtosis for symmetric initial
conditions.  The evolved PDF's for these two models are given in figure
7 (bilateral distribution) and figure 8 (uniform distribution).
In both cases, the singularities in the initial density distribution
remain in the evolved PDF.  However, despite the extreme nature of the
initial distribution functions, in both cases the evolved PDF
shows the expected qualititative behavior, with increasing skewness
and the development of a tail at large $\eta$.

\vskip 1 cm

\noindent{\bf 5 CONCLUSIONS}

Despite the simple-mindedness of local Lagrangian approximations,
our ``exact" approximation
provides remarkable agreement with numerical simulations of the evolution
of the density distribution function
with Gaussian initial conditions.  This agreement can be understood
in terms of the fact that this approximation reproduces nearly exactly
the hierarchical amplitudes at tree level.
The planar and spherical approximations appear to bound the evolution
of the PDF but are much less useful (unless, of course,
one is interested in the evolution of one-dimensional density fields,
in which case the planar approximation is exact for any initial conditions).

For the case of non-Gaussian initial conditions, we cannot be as confident.
Unlike the case of Gaussian initial conditions, the evolution of
the hierarchical amplitudes is non-local, as has been shown by
Fry \& Scherrer (1994) and Chodorowski \& Bouchet (1996).
Thus, no local approximation can exactly reproduce the hierarchial
amplitudes for non-Gaussian initial conditions.  However, the exact approximation does
a reasonable job of reproducing the hierarchical amplitudes for
limiting cases where a local approximation is valid.  Furthermore,
the application of this approximation to various non-Gaussian initial
conditions does show reasonable agreement with the expected qualitative behavior.

The mapping which gives us the exact approximation can also be applied
backwards, to map the evolved distribution function back onto the initial
distribution function.  This procedure is guaranteed to produce
the correct initial distribution function only for the case of Gaussian
initial conditions, and only for reasonably small $\sigma_l$, but
it is obvious from Fig. 1a that it would be highly accurate in this case.
This method should, in principle, be capable of distinguishing Gaussian
from non-Gaussian initial conditions, even if it could not accurately
give the exact form of the latter.

\vskip 0.5in

We thank Josh Frieman for helpful discussions.  We are grateful to David Weinberg
for providing us with results of his gravitational clustering simulations.
R.J.S. was supported in part by the
Department of Energy (DE-AC02-76ER01545).  Z.A.M.P. and R.J.S were
supported in part
by NASA (NAG 5-2864).

\vskip 1 cm

\begin{appendix}

\centerline{\bf APPENDIX:  The Vertex Generating Function for Local Approximations}

Here we demonstrate the simple form of the vertex generating function for
local Lagrangian approximations applied to Gaussian initial conditions;
namely, for a local Lagrangian mapping
$\eta = N(t)f(\delta_l)$, where $N(t)$ is a function only of $\sigma_l$,
the vertex generating function is given by
\begin{equation}
\label{AGdelta}
G_\delta(\tau) = f(\delta_l) - 1
\end{equation}
Our argument proceeds in two stages.  First we show that this relation
holds for the simpler case of local Eulerian mappings.  Then we show
that the vertex generating function is the same for local Eulerian
and local Lagrangian mappings.

Consider first the local Eulerian approximation
\begin{equation}
\label{eulerian}
\eta(\xvec) = N_E(t) f(\delta_l(\xvec)).
\end{equation}
The mapping $f$ can be expanded in a power series
\begin{equation}
f(\delta_l) = \sum_{j=0}^\infty b_j \delta_l^j
\end{equation}
where $b_0 = 1$,
and $N(t)$, for Gaussian initial conditions, can be a function only
of $\sigma_l$:
\begin{equation}
N_E(t) = \sum_{k=0}^\infty c_k \sigma_l^k.
\end{equation}
where $c_0 = 1$.
These three equations give us the expression for the $n^{th}$ order
expansion of $\delta$:
\begin{equation}
\delta^{(n)} = \sum_{j+k=n} b_j c_k \delta_l^j \sigma^k.
\end{equation}
We can substitute this expression into equation (\ref{vertex2}) to
obtain $\nu_n$.  When we do this and take the connected average, all
of the terms vanish except for the term $j=n$, $k=0$; the other
terms are not connected.  The final result is
\begin{eqnarray}
\nu_n &=& {\int \langle b_n \delta^{(1)^n}(\xvec) \delta^{(1)}(\xvec_1)...\delta^{(1)}(\xvec_n)\rangle_c
d^3\xvec ~ d^3 \xvec_1...d^3\xvec_n
\over(\int \langle \delta^{(1)}(\xvec)\delta^{(1)}(\xvec^\prime) \rangle
d^3\xvec ~ d^3 \xvec^\prime)^n} \nonumber\\
 &=& b_n n!
\end{eqnarray}
since there are $n!$ different connected graphs.
Then from equation (\ref{vertex1}) we obtain
\begin{eqnarray}
G_\delta(\tau) &=& \sum_{n=1}^\infty b_n \tau^n \nonumber\\
&=& f(\tau) - 1
\end{eqnarray}

Thus, equation (\ref{Gdelta}) holds for local Eulerian mappings; we
now show that $G_\delta(\tau)$ is the same if we take $f(\delta_l)$ to
be a local Lagrangian mapping instead of a local Eulerian mapping.
To do this, we first note that $G_\delta$ is completely determined
by the values of $\nu_n$ (equation \ref{vertex1}), which are, in turn,
completely determined by the $S_p(0)$'s (equation \ref{Snu}).  Thus, it suffices to show that
the local Eulerian mapping given by equation (\ref{eulerian})
and the local Lagrangian mapping
\begin{equation}
\label{lagrangian}
\eta(\qvec) = N_L(t) f(\delta_l(\qvec))
\end{equation}
have the same values for $S_p(0)$, which is equivalent to the statement
that they have identical cumulants $\kappa_p$ in the limit $\sigma \rightarrow 0$.

To demonstrate this, we introduce the
characteristic function $\phi_E(t)$ for the Eulerian mapping, which is the Fourier transform of
the PDF:
\begin{equation}
\phi_E(t) = \int P_E(\eta) e^{i\eta t} d\eta
\end{equation}
where we assume that $P_E(\eta)$ is the PDF for the Eulerian mapping given by
equation (\ref{eulerian}).  If instead we use the same function to produce
the local Lagrangian mapping given by equation (\ref{lagrangian}), we
obtain the PDF $P_L(\eta)$, with corresponding characteristic
function $\phi_L(t)$.
Using
equation (\ref{p2value}) for the Lagrangian mapping, and its
equivalent (without the $1/\eta$ factor) for the Eulerian mapping, we find
that $\phi_L$ and $\phi_E$ are related by:
\begin{equation}
\label{phiEL}
\phi_E({N_L \over N_E} t) = {1\over i} \phi_L^\prime(t),
\end{equation}
where $N_L$ and $N_E$ are the normalizing factors for the Lagrangian
and Eulerian mappings, given respectively by
\begin{equation}
N_L = \langle {1\over f(\delta_l)} \rangle
\end{equation}
and
\begin{equation}
N_E = {1\over \langle f(\delta_l) \rangle}.
\end{equation}
The cumulants we wish to calculate are related to the characteristic
function by
\begin{equation}
\label{characteristic}
\phi(t) = \exp(\sum_{p=1}^\infty {(it)^p \over p!} \kappa_p)
\end{equation}
If we let $\kappa^{(E)}_p$ represent the cumulants for the Eulerian
mapping, and $\kappa^{(L)}_p$ represent the cumulants for the Lagrangian
mapping, then we can substitute equation (\ref{characteristic}) into
equation (\ref{phiEL}) to obtain:
\begin{equation}
\label{log}
\sum_{p = 1}^ \infty [\kappa^{(E)}_p(\langle f \rangle \langle 1/f \rangle)^p - \kappa^{(L)}_p] {(it)^p \over p!}
= \ln[1+ \sum_{p=1}^\infty \kappa^{(L)}_{p+1} {(it)^p \over p!}],
\end{equation}
where we have used the fact that $\kappa^{(L)}_1 = 1$.  Now
we note that for both the Eulerian and Lagrangian mappings considered here,
the ``evolved" distribution is hierarchical, so that
$\kappa_p^{(L)} \sim O(\sigma^{2(p-1)})$, and $\kappa_p^{(E)} \sim O(\sigma^{2(p-1)})$.
Furthermore, $(\langle f \rangle \langle 1/f \rangle)^p = 1+ O(\sigma^2).$
Expanding out the right-hand side of equation (\ref{log}) and equating
terms with equal powers of $t^p$, we find that
\begin{equation}
\kappa^{(E)}_p - \kappa^{(L)}_p \sim O(\sigma^{2p}).
\end{equation}
Furthermore, $\kappa_p$ for $P(\eta)$ and $\kappa_p$ for $P(\delta)$
are identical for $p>1$, since $\eta$ and $\delta$ differ by a constant.
Thus, the difference between $S_p(\sigma)$ for the Lagrangian local
mapping and $S_p(\sigma)$ for the local Eulerian mapping vanishes in the limit
where $\sigma \rightarrow 0$, so that the two mappings have the same
$G_\delta(\tau)$.

These results allow us to generate, in a simple way, two different
mappings which, when
applied to Gaussian initial conditions, yield a density field with any desired
vertex generating function $G_\delta(\tau)$ or hierarchical amplitudes $S_p(0)$.

\end{appendix}

\vfill
\eject
\centerline{\bf REFERENCES} 

\vskip 1 cm 

\noindent Bernardeau, F., 1992, ApJ, 392, 1

\noindent Bernardeau, F., 1994, A \& A, 291, 697

\noindent Bernardeau, F., \& Kofman, L., 1995, ApJ, 443, 479

\noindent Betancort-Rijo, J., 1991, MNRAS, 251, 399

\noindent Chodorowski, M.J., \& Bouchet, F.R., 1996, MNRAS, in press

\noindent Coles, P., Moscardini, L., Lucchin, F., Matarrese, S., \& Messina,
A. 1993, MNRAS, 264,

\indent 749

\noindent Fry, J.N., 1984, ApJ, 279, 499

\noindent Fry, J.N., \& Scherrer, R.J., 1994, ApJ, 429, 36

\noindent Juszkiewicz, R., Weinberg, D.H., Amsterdamski, P., Chodorowski, M.,
\& Bouchet, F.,

\indent 1995, ApJ, 442, 39

\noindent Kofman, L., 1991, in Primordial Nucleosynthesis and Evolution of Early
Universe,

\indent Ed. K. Sato \& J. Audouze (Dordrecht:  Kluwer), 495

\noindent Kofman, L., Bertschinger, E., Gelb, J.M., Nusser, A., \& Dekel, A.,
1994, ApJ, 420, 44

\noindent Lokas, E.L., Juszkiewicz, R., Bouchet, F.R., \& Hivon, E. 1996,
ApJ, in press

\noindent Lokas, E.L., Juszkiewicz, R., Weinberg, D.H., \& Bouchet, F.R.
1995, MNRAS, 274, 730

\noindent Matarrese, S., Lucchin, F., Messina, A., \& Moscardini L. 1991, 
MNRAS, 253, 35

\noindent Matarrese, S., Lucchin, F., Moscardini, L., \& Saez, D. 1992,
MNRAS, 259, 437

\noindent Messina, A., Moscardini, L., Lucchin, F., \& Matarrese, S. 1990, 
MNRAS,  245, 244

\noindent  Moscardini, L., Matarrese, S., Lucchin, F., \& Messina A. 1991, 
MNRAS,  248, 424

\noindent Munshi, D., Sahni, V., \& Starobinsky, A.A. 1994, ApJ, 436, 517

\noindent Nusser, A., Dekel, A., Bertschinger, E., \& Blumenthal, G.R. 1991,
ApJ, 379, 6

\noindent Padmanabhan, T., \& Subramanian, K. 1993, ApJ, 410, 482

\noindent Peebles, P.J.E. 1980, The Large-Scale Structure of the Universe
(Princeton University Press)

\noindent Sahni, V., \& Coles, P. 1995, Phys. Rep., 262, 1

\noindent Scoccimarro, R. \& Frieman, J. 1996, ApJ, submitted

\noindent Shandarin, S.F., \& Zel'dovich, Ya. B. 1989, Rev Mod Phys, 61, 185

\noindent Stuart, A., \& Ord, J.K. 1987, Kendall's Advanced Theory of Statistics,
Vol. 1 (London:

\indent Charles Griffin)

\noindent Weinberg, D.H., \& Cole, S. 1992, MNRAS,  259, 652

\noindent Zel'dovich, Ya.B. 1970, A \& A, 5, 84

\vfill
\eject

\centerline{\bf FIGURE CAPTIONS}

\vskip 0.5 cm

\noindent Figure 1:  Comparison of our exact approximation for
the evolution of the smoothed density distribution function (solid
curve)
with the results of a numerical simulation of gravitational clustering
(points with error bars),
for Gaussian initial conditions with 
power spectrum $P(k) \propto k^{-1}$ and spherical
top-hat smoothing.  The distribution functions are calculated
for a linearly evolved rms fluctuation
of (a)  $\sigma_l = 0.5$, (b) $\sigma_l = 0.75$, (c) $\sigma_l = 1.0$.

\vskip 0.5 cm

\noindent Figure 2:  As Figure 1, but here the
results of the numerical simulation of gravitational
clustering (points with error bars) are compared with the
smoothed density distributions given by the
planar approximation (dotted curve)
and the spherical approximation (dashed curve) for
(a) $\sigma_l = 0.5$,
(b) $\sigma_l = 0.75$.

\vskip 0.5 cm

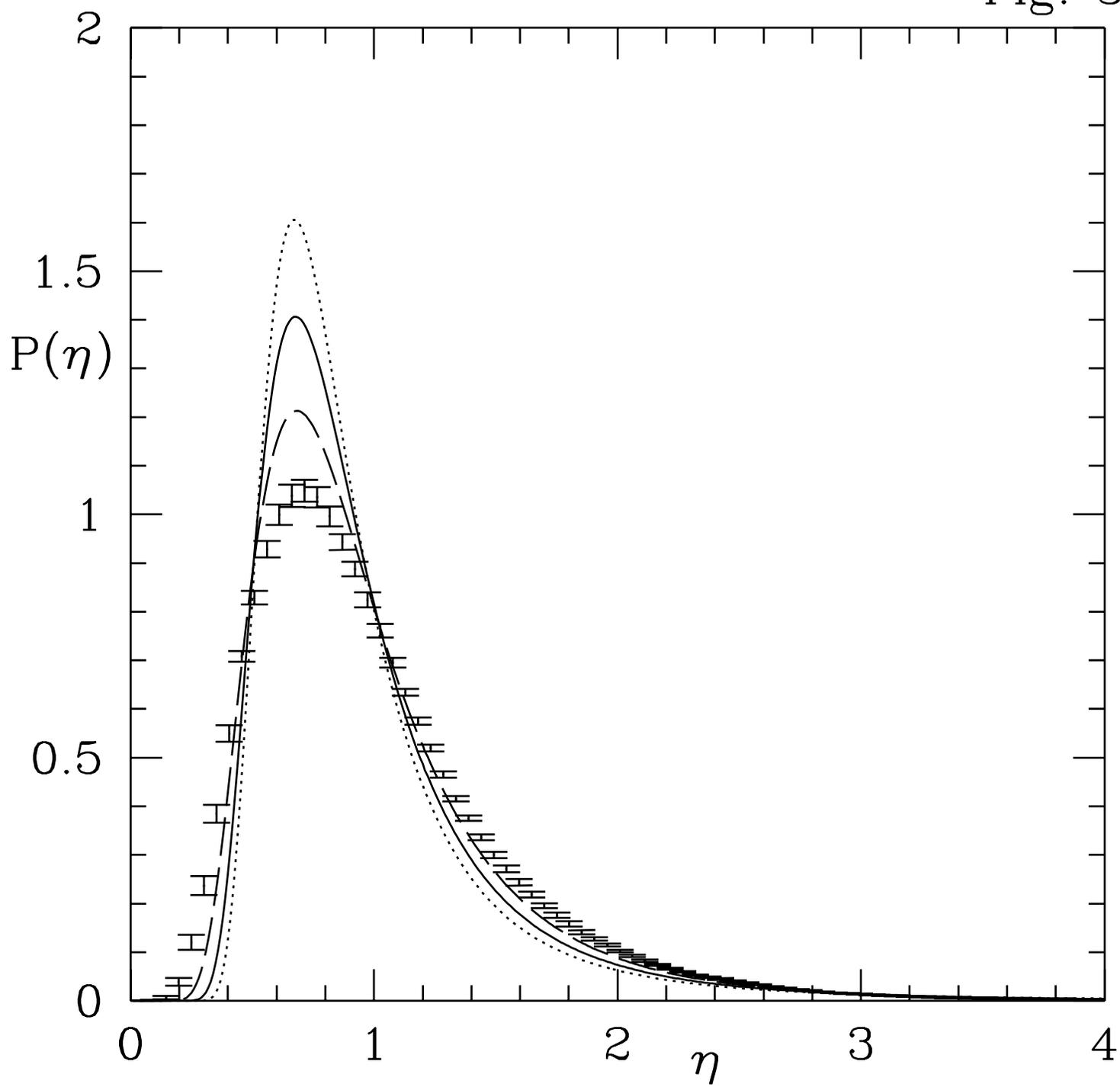
\noindent Figure 3:  As Figure 1, but here the results
of the numerical simulation of gravitational clustering (points
with error bars) are compared
with the unsmoothed density distributions given by
the exact approximation (solid curve),
the planar approximation (dotted curve), and the spherical approximation (dashed curve), for $\sigma_l = 0.5$.

\vskip 0.5 cm

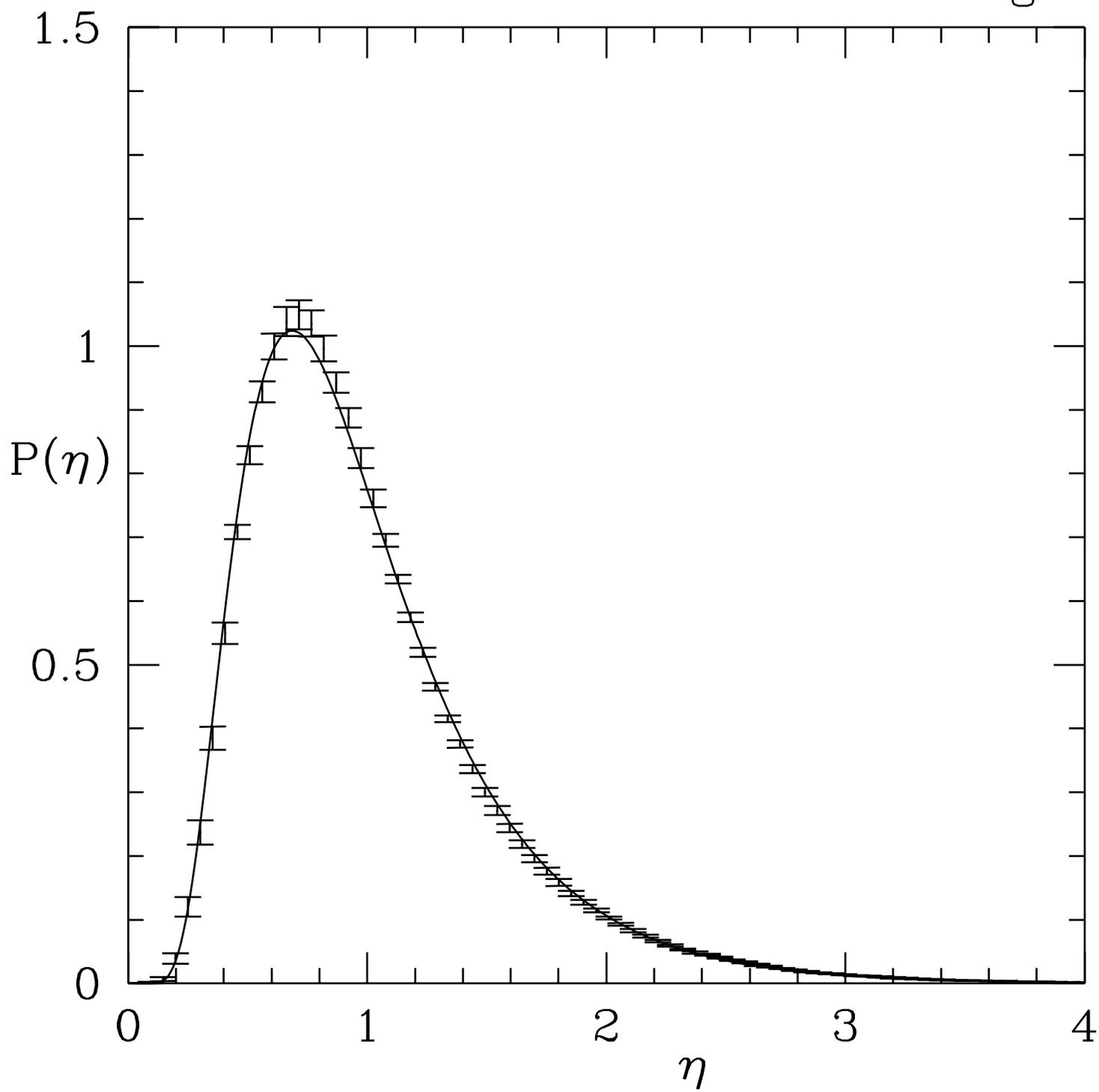
\noindent Figure 4:  As Figure 1a, using an Eulerian version of the exact approximation (solid curve).

\vskip 0.5 cm

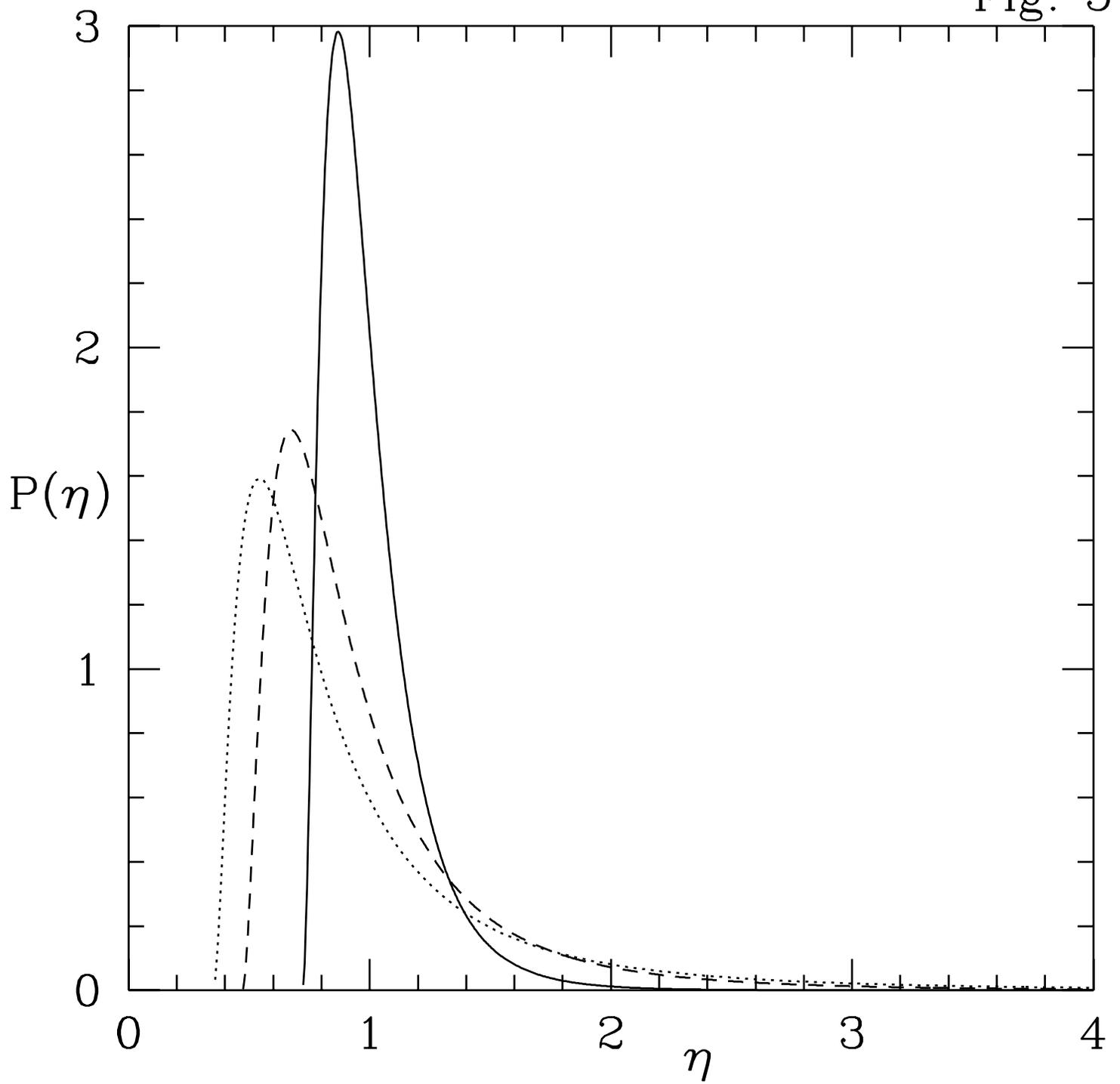
\noindent Figure 5:  Evolved smoothed density distribution functions
given by the exact approximation
with $P(k) \propto k^{-1}$, for an initial gamma function
density distribution
with positive skewness.  The distribution functions are calculated
for a linearly evolved rms fluctuation of $\sigma_l = 0.2$
(solid curve), $\sigma_l = 0.5$ (dashed curve), and
$\sigma_l = 0.75$ (dotted curve).

\vskip 0.5 cm

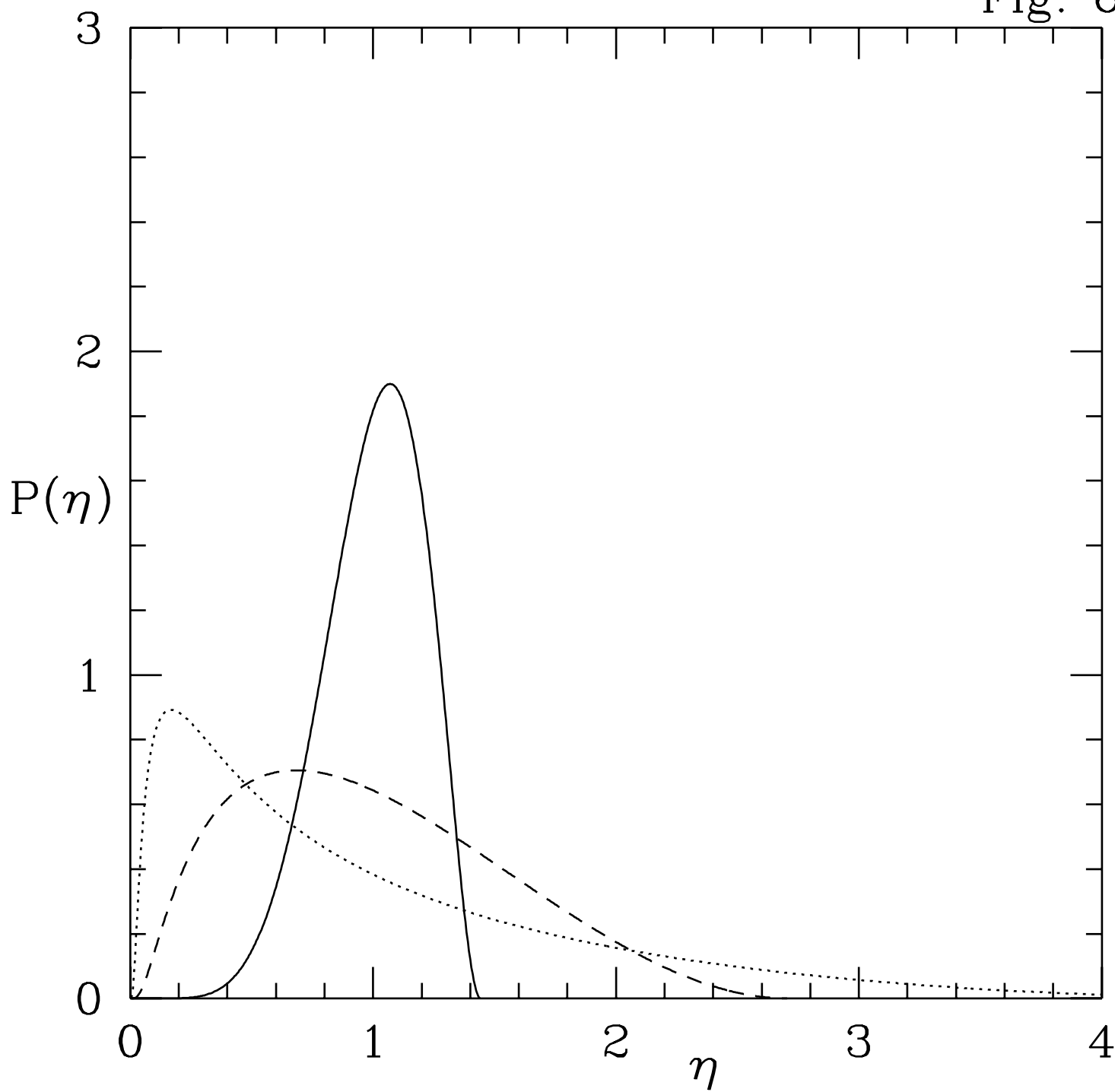
\noindent Figure 6:  As Figure 5, for an initial gamma function with negative skewness.

\vskip 0.5 cm

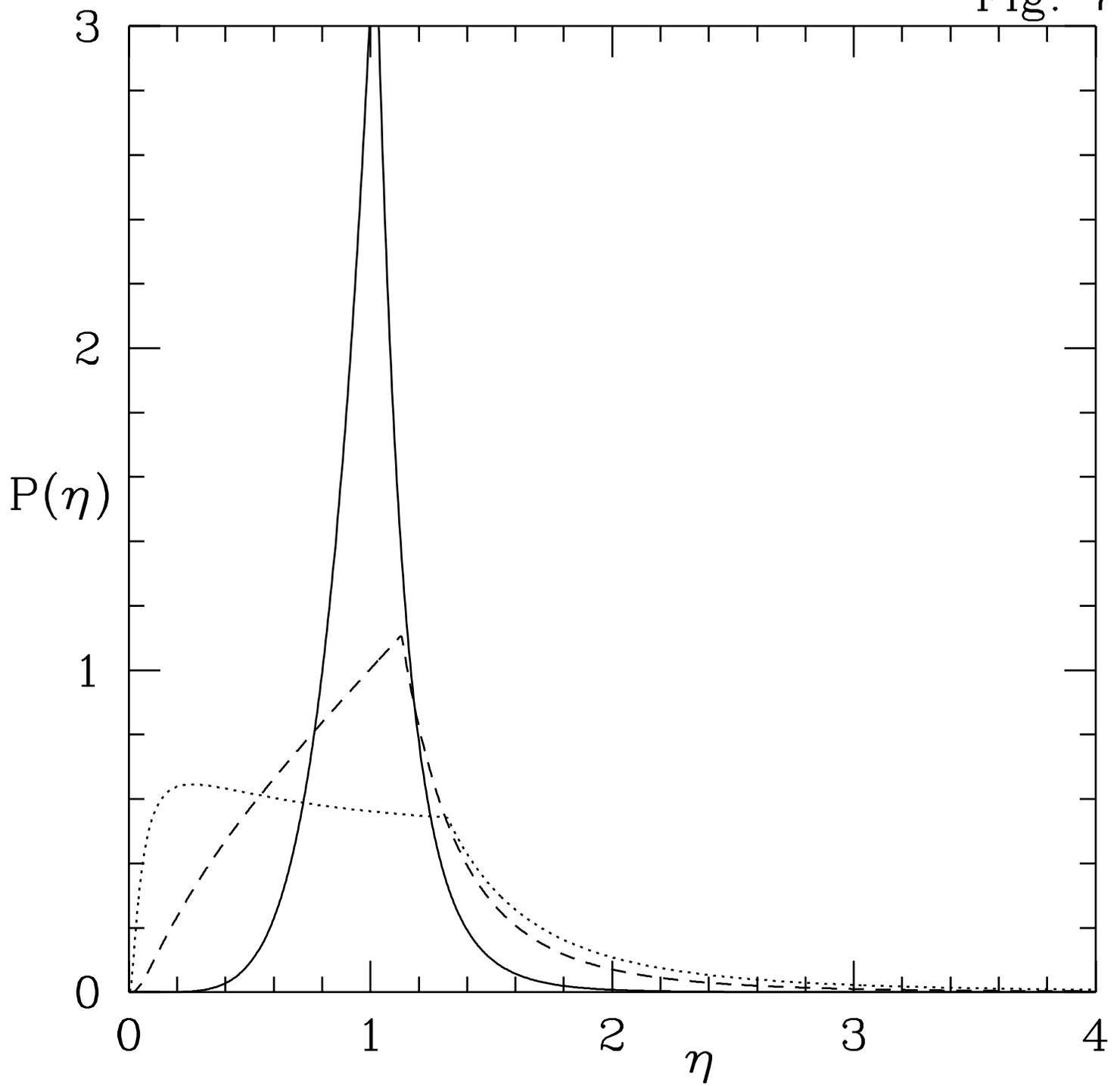
\noindent Figure 7:  As Figure 5, for an initial bilateral exponential function.

\vskip 0.5 cm

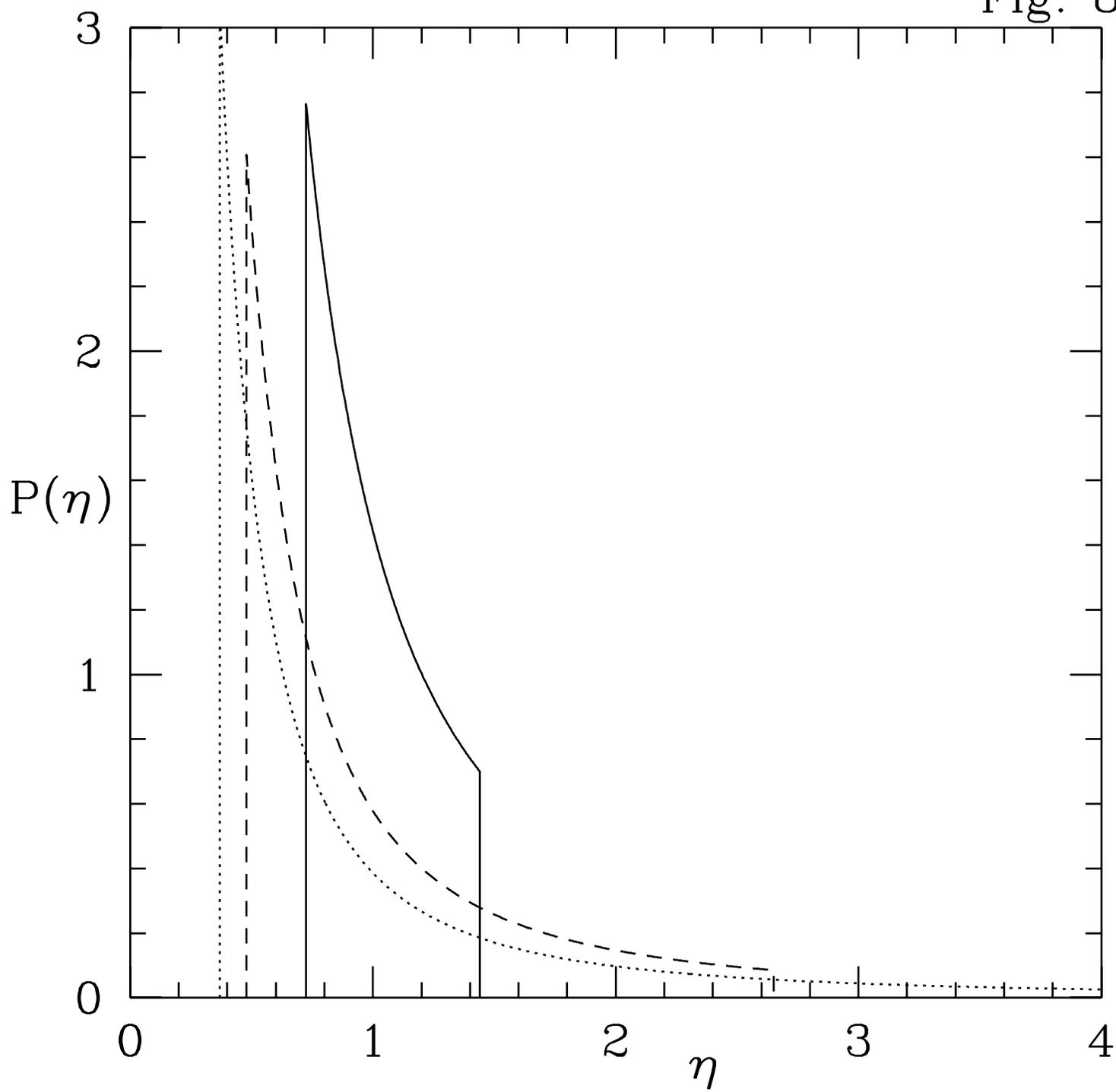
\noindent Figure 8:  As Figure 5, for an initial uniform distribution.

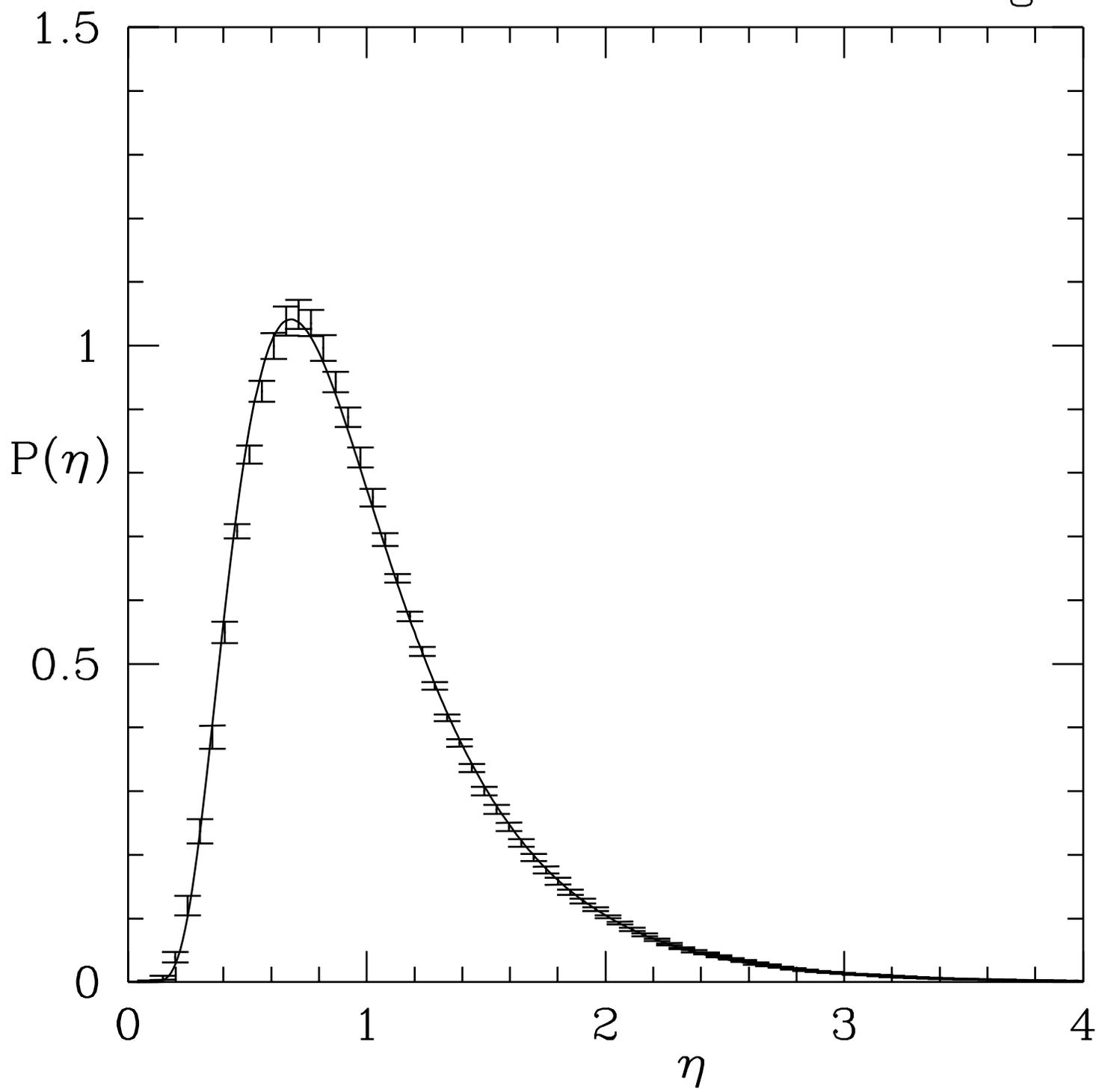
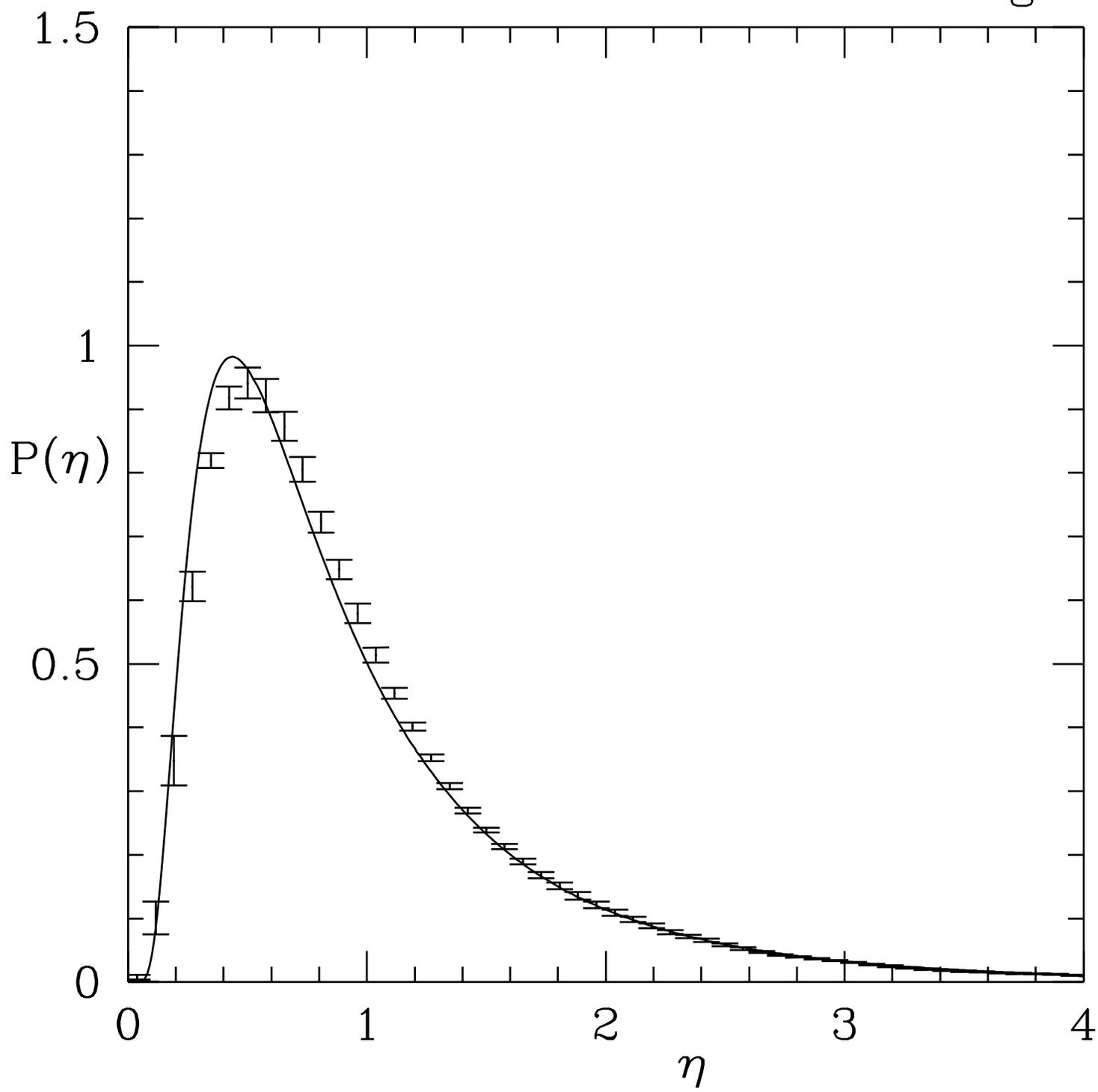
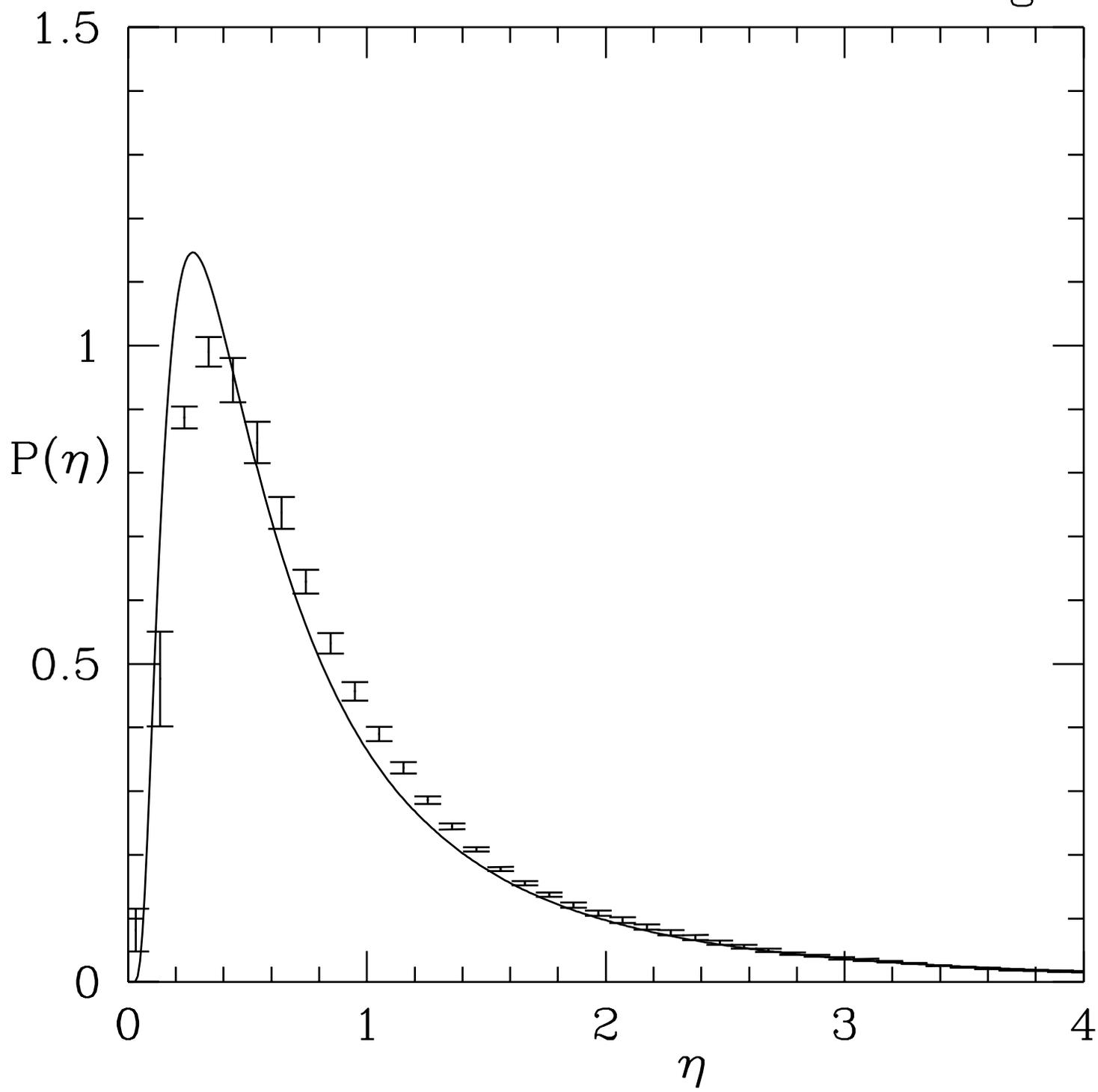
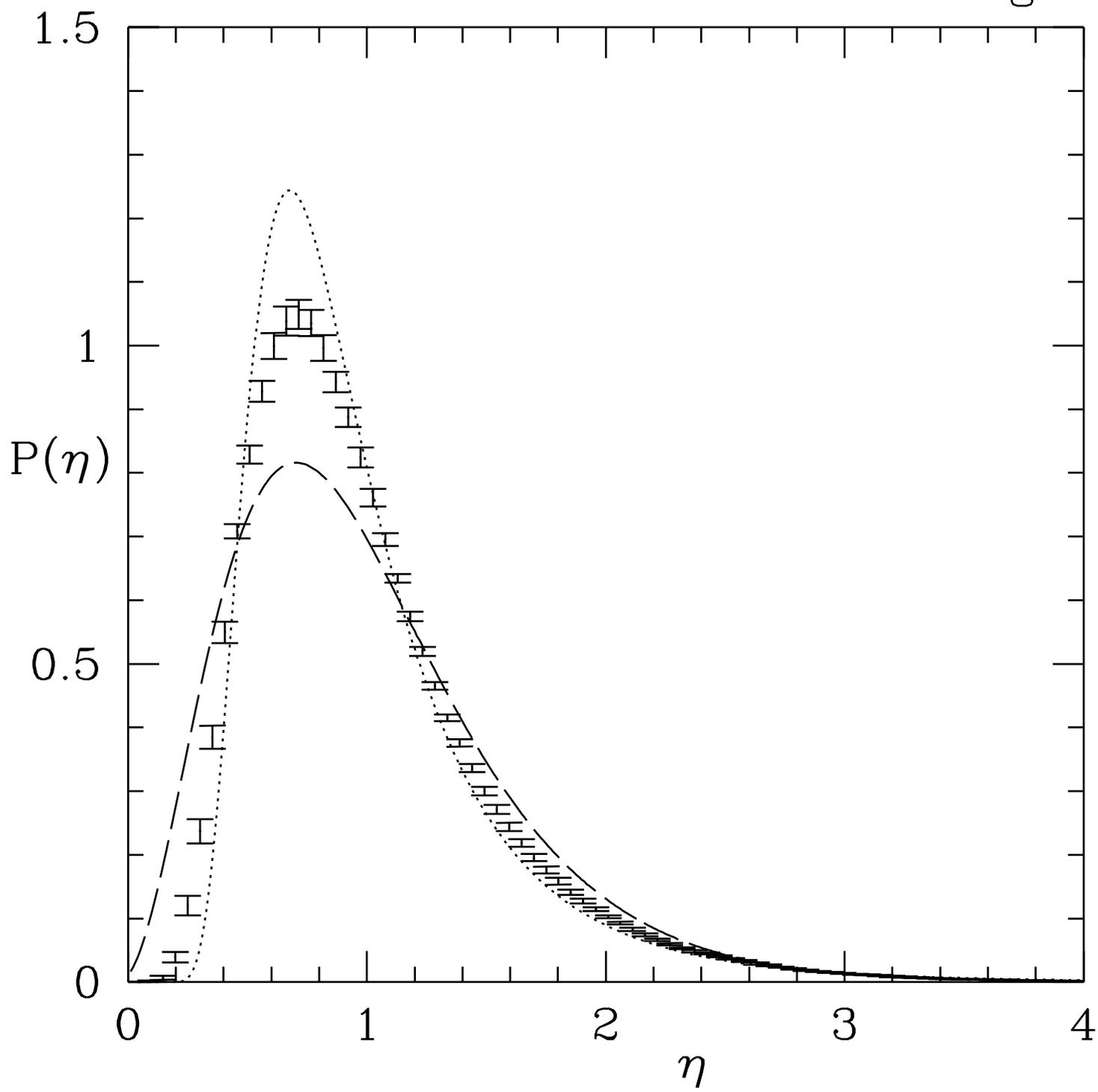
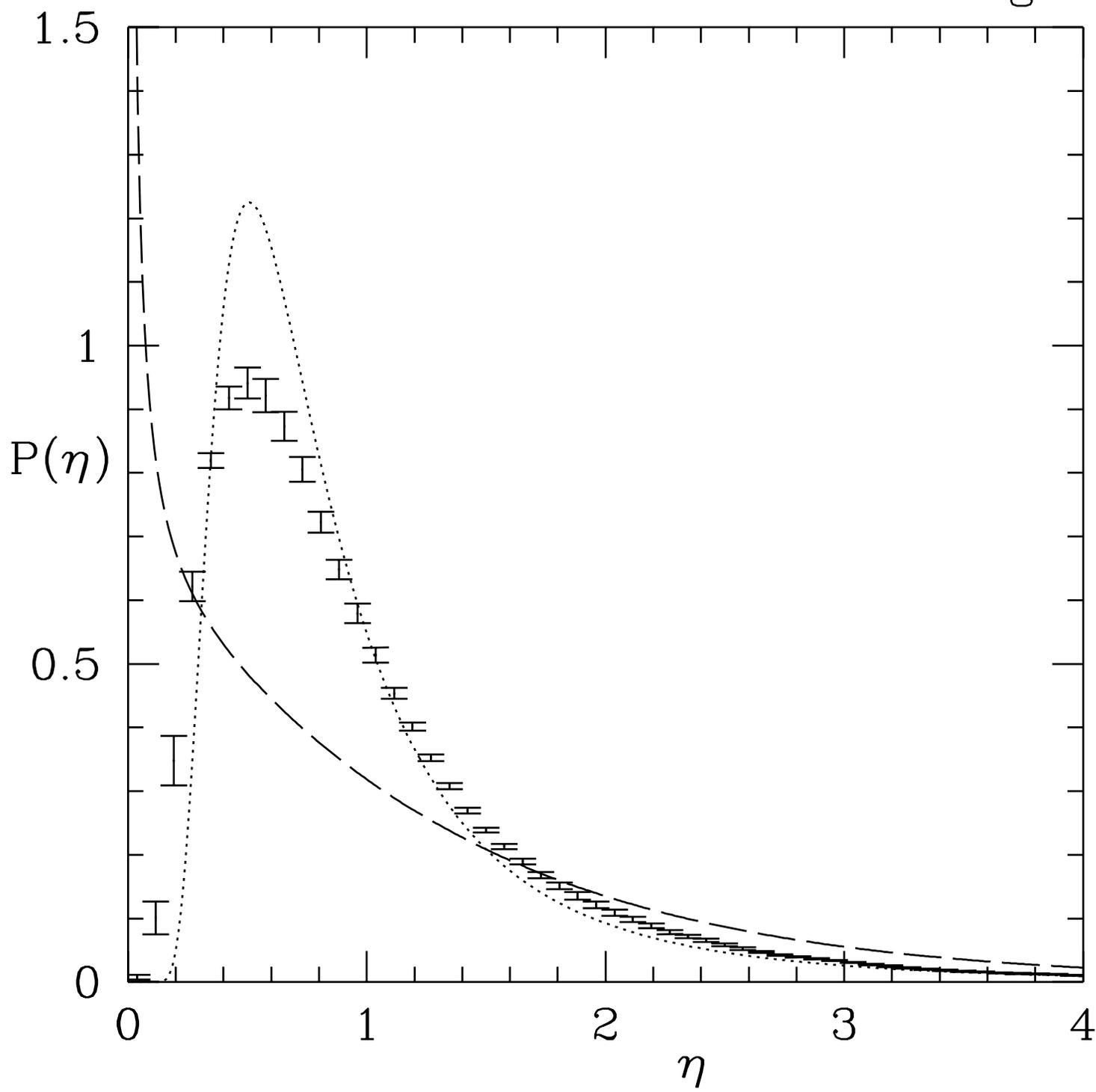
\end{document}